\documentclass[12pt]{article}
\usepackage{axodraw,cite,epsfig}

\newcommand{\appendixA}{\setcounter{equation}{0}
\def\theequation{\rm{A}.\arabic{equation}}\section*}
\newcommand{\appendixB}{\setcounter{equation}{0}
\def\theequation{\rm{B}.\arabic{equation}}\section*}

\catcode`\@=11
\def\marginnote#1{}
\def\ifmath#1{\relax\ifmmode #1\else $#1$\fi}

\def\hl{h^0}
\def\mhl{m_{\hl}}

\def\mstop{m_{\,\widetilde{t}}}
\def\stop{\tilde{t}}

\def\bold#1{\setbox0=\hbox{$#1$}%
     \kern-.025em\copy0\kern-\wd0
     \kern.05em\copy0\kern-\wd0
     \kern-.025em\raise.0433em\box0 }
\def\eighth{\ifmath{{\textstyle{1 \over 8}}}}
\def\GENITEM#1;#2{\par\vskip6pt \hangafter=0 \hangindent=#1
   \Textindent{$ #2$ }\ignorespaces}

\def\draftlabel#1{{\@bsphack\if@filesw {\let\thepage\relax
  \xdef\@gtempa{\write\@auxout{\string
    \newlabel{#1}{{\@currentlabel}{\thepage}}}}}\@gtempa
    \if@nobreak \ifvmode\nobreak\fi\fi\fi\@esphack}
     \gdef\@eqnlabel{#1}}
\def\@eqnlabel{}
\def\@vacuum{}
\def\draftmarginnote#1{\marginpar{\raggedright\scriptsize\tt#1}}
\def\draft{\oddsidemargin -.5truein
        \def\@oddfoot{\sl preliminary draft \hfil
        \rm\thepage\hfil\sl\today\quad\militarytime}
        \let\@evenfoot\@oddfoot \overfullrule 3pt
        \let\label=\draftlabel
        \let\marginnote=\draftmarginnote

\def\@eqnnum{(\theequation)\rlap{\kern\marginparsep\tt\@eqnlabel}%
\global\let\@eqnlabel\@vacuum}  }
\def\preprint{\twocolumn\sloppy\flushbottom\parindent 1em
        \leftmargini 2em\leftmarginv .5em\leftmarginvi .5em
        \oddsidemargin -.5in    \evensidemargin -.5in
        \columnsep 15mm \footheight 0pt
        \textwidth 250mmin      \topmargin  -.4in
        \headheight 12pt \topskip .4in
        \textheight 175mm
        \footskip 0pt

\def\@oddhead{\thepage\hfil\addtocounter{page}{1}\thepage}
        \let\@evenhead\@oddhead \def\@oddfoot{} \def\@evenfoot{}
}
\def\titlepage{\@restonecolfalse\if@twocolumn\@restonecoltrue\onecolumn
     \else \newpage \fi \thispagestyle{empty}\c@page\z@
        \def\thefootnote{\fnsymbol{footnote}} }
\def\endtitlepage{\if@restonecol\twocolumn \else  \fi
        \def\thefootnote{\arabic{footnote}}
        \setcounter{footnote}{0}}  %\c@footnote\z@ }
\catcode`@=12
\relax
\newlength{\captsize}          \let\captsize=\footnotesize
\newlength{\captwidth}         
\newlength{\beforetableskip}   
\newcommand{\capt}[1]{\begin{minipage}{\captwidth}
               \let\normalsize=\captsize
               \caption[#1]{#1}
               \end{minipage}\\  \vspace{\beforetableskip}}
\makeatletter
\newcommand{\fcaption}[1]{
         \refstepcounter{figure}
         \setbox\@tempboxa = \hbox{\footnotesize Figure~\thefigure. #1}
         \ifdim \wd\@tempboxa > 12cm
             {\footnotesize \textbf{Figure~\thefigure.} #1}
         \else
              {\begin{center}
              {\footnotesize \textbf{Figure~\thefigure.} #1}
               \end{center}}
         \fi}
\makeatother
\newenvironment{Eqnarray}%
     {\arraycolsep 0.14em\begin{eqnarray}}{\end{eqnarray}}
\def\be{\begin{equation}}
\def\ee{\end{equation}}
\def\bea{\begin{Eqnarray}}
\def\eea{\end{Eqnarray}}

\def\eq#1{eq.~(\ref{#1})}
\def\fig#1{fig.~\ref{#1}}

\def\Ref#1{ref.~\cite{#1}}
\def\Refs#1#2{refs.~\cite{#1} and \cite{#2}}

\def\eqs#1#2{eqs.~(\ref{#1})--(\ref{#2})}
\def\Eq#1{Eq.~(\ref{#1})}

\def\eqns#1#2{eqs.~(\ref{#1}) and (\ref{#2})}
\def\tanb{\tan\beta}
\def\cotb{\cot\beta}
\def\hl{h}
\def\ha{A}
\def\hh{H}

\def\mha{m_{\ha}}
\def\mhl{m_{\hl}}
\def\mhh{m_{\hh}}
\def\mz{m_Z}

\def\mst11{m_{\;\widetilde{t}_{1}}}

\def\mst22{m_{\;\widetilde{t}_{2}}}
\def\mst12{m_{\;\widetilde{t}_{1,2}}}

\def\msb11{m_{\;\widetilde{b}_{1}}}
\def\msb22{m_{\;\widetilde{b}_{2}}}
\def\msb12{m_{\;\widetilde{b}_{1,2}}}

\def\mwidetilde2{\widetilde{m}^{2}}

\def\leff{\lambda}
\def\mbart{\overline{m}_{t}}

\def\nicefrac#1#2{\hbox{${#1\over #2}$}}
\def\quarter{\nicefrac{1}{4}}
\def\half{\nicefrac{1}{2}}
\def\third{\nicefrac{1}{3}}
\def\eighth{\nicefrac{1}{8}}
\def\lsim{\mathrel{\raise.3ex\hbox{$<$\kern-.75em\lower1ex\hbox{$\sim$}}}}
\def\gsim{\mathrel{\raise.3ex\hbox{$>$\kern-.75em\lower1ex\hbox{$\sim$}}}}
\def\drbar{$\overline{\rm DR}$\ }
\def\stopa{\widetilde t_1}
\def\stopb{\widetilde t_2}
\def\mstopa{m_{\tilde t_1}}
\def\mstopb{m_{\tilde t_2}}
\def\mstop{m_{\tilde t_j}}
\def\msusy{M_{\rm SUSY}}
\def\pslash{\not{\hbox{\kern-2.3pt $p$}}}
\def\Azero{{\cal A}_0}
\def\Bzero{{\cal B}_0}
\def\Bone{{\cal B}_1}
\def\Bn{{\cal B}_n}
\def\ifmath#1{\relax\ifmmode #1\else $#1$\fi}
\def\ls#1{\ifmath{_{\lower1.5pt\hbox{$\scriptstyle #1$}}}}
\def\lsup#1{\ifmath{^{\lower2pt\hbox{$\scriptstyle #1$}}}}
\def\gammav{\gamma\ls{5}}
\def\Xtbarii{\overline{X}_t\lsup{\,2}}
\def\Xtbariii{\overline{X}_t\lsup{\,3}}
\def\Xtbariv{\overline{X}_t\lsup{\,4}}
\def\Msbarii{\overline{M}_S\lsup{\,2}}
\def\Msbariii{\overline{M}_S\lsup{\,3}}
\def\Msbariv{\overline{M}_S\lsup{\,4}}
\def\Xtbarp{\overline{X}_t\lsup{\,\prime}}
\relax

\newcommand{\msbar}{$\overline{\mathrm{MS}}$}
\newcommand{\MS}{\overline{\mathrm{MS}}}

\newcommand{\oas}{{\cal O}(\alpha_s)}
\newcommand{\oaas}{{\cal O}(\alpha\alpha_s)}
\newcommand{\cp}{{\cal CP}}

\newcommand{\twol}{two-loop}
\newcommand{\onel}{one-loop}
\newcommand{\mma}{{\em Mathematica}}

\newcommand{\fh}{{\em FeynHiggs}}

\newcommand{\mh}{m_h}

\newcommand{\msq}{M_{\mathrm{SUSY}}}
\newcommand{\MstL}{M_{\tilde{t}_L}}
\newcommand{\MstR}{M_{\tilde{t}_R}}
\newcommand{\ms}{M_S}
\newcommand{\Msbar}{\overline{M}_S}
\newcommand{\Xt}{X_t}
\newcommand{\Xtbar}{\overline{X}_t}

\newcommand{\mt}{m_t}
\newcommand{\mtbar}{\overline{m}_{t}}

\newcommand{\mtsmreg}{m^{\MS}_{t, \mathrm{SM}}}

\newcommand{\tst}{\theta_{\tilde{t}}}
\newcommand{\mste}{m_{\tilde{t}_1}}
\newcommand{\mstz}{m_{\tilde{t}_2}}
\newcommand{\Stop}{\widetilde{t}}
\newcommand{\dst}{\Delta_{\tilde{t}}}
\newcommand{\mgl}{m_{\tilde{g}}}

\newcommand{\os}{on-shell}
\newcommand{\OS}{\mathrm{OS}}
\newcommand{\de}{\delta}
\newcommand{\De}{\Delta}
\newcommand{\non}{\nonumber}

\begin{document}
\topmargin-1.5cm
%\draft
%\preprint
%
\begin{titlepage}
\begin{flushright}
ANL-HEP-PR-99-109\\
CERN-TH/2000-005\\
DESY 99-197\\
FERMILAB-Pub-00/028-T\\
KA-TP-10-1999\\
SCIPP-99/46\\
hep--ph/0001002 \\
\end{flushright}
\vskip 0.1in
\begin{center}
{\Large\bf  Reconciling the Two-Loop Diagrammatic and}\\[0.5em]
{\Large\bf  Effective Field Theory Computations of the Mass}\\[0.5em]
{\Large\bf  of the Lightest $\cp$-even Higgs Boson in the MSSM}
%\footnote{Work supported in part by the European Union
%(contract CHRX/CT92-0004) and CICYT of Spain
%(contract AEN95-0195).}
\vskip 0.3in
{\bf M. Carena~$^{\S,\dagger}$},
{\bf H.E. Haber~$^{\sharp}$},
{\bf S. Heinemeyer~$^{\ddagger}$},\\
~\\
{\bf W. Hollik~$^{\P}$},
{\bf C.E.M. Wagner~$^{\dagger,\ast,\natural}$}
{\bf and} {\bf G. Weiglein~$^{\dagger}$}
\vskip0.2in
$^{\S}$~FERMILAB, Batavia, IL 60510-0500 USA\\[.2em]
$^{\dagger}$~CERN, TH Division, CH--1211 Geneva 23,
Switzerland\\[.2em]
$^{\sharp}$~Santa Cruz Inst. for Part. Phys.,
Univ.~of California, Santa Cruz, CA 95064 USA\\[.2em]
$^{\ddagger}$~DESY Theorie, Notkestrasse 85, 22603 Hamburg,
Germany\\[.2em]
$^{\P}$~Institut f\"ur Theoretische Physik, Univ.~of
Karlsruhe, 76128 Karlsruhe, Germany\\[.2em]
$^{\ast}$~High
Energy Physics Division, Argonne National Lab., Argonne, IL 60439
USA\\[.2em] $^{\natural}
$~Enrico Fermi Institute, Univ.~of
Chicago, 5640 Ellis, Chicago, IL 60637 USA\\[2em]
\end{center}
%\vskip0.2cm
\begin{center}
{\bf Abstract}
\end{center}
%\begin{quote}
The mass of the lightest $\cp$-even Higgs boson of the minimal
supersymmetric extension of the Standard
Model (MSSM) has previously been computed including
${\cal O}(\alpha\alpha_s)$ two-loop contributions
by an on-shell diagrammatic method,
while approximate analytic results have also been obtained via
renormalization-group-improved effective potential and effective field
theory techniques.  Initial comparisons of the corresponding two-loop
results revealed an apparent discrepancy between terms that depend
logarithmically on the supersymmetry-breaking scale, and
different dependences of the non-logarithmic terms on the squark
mixing parameter, $X_t$.  In this paper, we determine the origin of
these differences as a consequence of different renormalization schemes
in which both calculations are performed. By re-expressing the on-shell
result in terms of \msbar\ parameters, the
logarithmic two-loop contributions obtained by the different approaches
are shown to coincide.  The remaining difference, arising from genuine
non-logarithmic two-loop contributions, is identified, and its
%the origin of the different
%dependence on $X_t$ of the two-loop
%non-logarithmic contributions is clarified, and its
effect on the maximal value of the lightest $\cp$-even Higgs boson
mass is discussed.  Finally, we show that
in a simple analytic approximation to the Higgs mass, the leading
two-loop radiative corrections can be absorbed to a large extent into an
effective one-loop expression by evaluating the running top quark mass
at appropriately chosen energy scales.
%\end{quote}
%\vskip0.5cm
%\begin{flushleft}
%CERN-TH/96-242\\
%October 1999 \\
%\end{flushleft}

\end{titlepage}
\setcounter{footnote}{0}
\setcounter{page}{0}
\newpage
%
% BODY
\section{Introduction}

In the minimal supersymmetric extension of the Standard Model (MSSM),
the mass of the lightest $\cp$-even Higgs boson, $\mhl$, is calculable
as a function of the MSSM parameters.  At tree-level, $\mhl$ is a
function of the $\cp$-odd Higgs boson mass, $\mha$, and the ratio of
vacuum expectation values, $\tanb$.  Moreover, the tree-level value of
$\mhl$ is bounded by $\mhl\leq \mz|\cos 2\beta|$, which is on the verge
of being ruled out by the LEP Higgs search~\cite{lephiggs}.  When
radiative corrections
are taken into account, $\mhl$ depends in addition on the MSSM
parameters that enter via virtual loops.  The radiatively corrected
value of $\mhl$ depends most sensitively on the parameters of the
top-squark (stop) sector: the average squared-mass of the two stops,
$\ms^2$, and the off-diagonal stop squared-mass parameter, $m_t X_t$.
The stop mixing parameter is $X_t\equiv A_t-\mu\cotb$, where $A_t$ is
the coefficient of the soft-supersymmetry-breaking stop-Higgs boson
tri-linear interaction term and $\mu$ is the supersymmetric Higgs mass
parameter.  The radiatively-corrected value of $\mhl^2$ is enhanced
by a factor of $G_F\mt^4$ and grows logarithmically as
$\ms$ increases~\cite{hhprl}.  In particular, the upper bound for
$\mhl$ (which is achieved when $\mha\gg\mz$ and $\tanb\gg 1$)
is significantly increased beyond its tree-level upper bound of $\mz$.

The complete one-loop diagrammatic computation of $\mhl$ has been
carried out in refs.~\cite{fulloneloop1,mhiggsOL2,fulloneloop2}.
However, for $\ms\gg\mt$, the logarithmically enhanced terms are
significant (in particular, the most significant logarithmic terms
are those that are enhanced by the $G_F m_t^4$ pre-factor noted above),
in which case
leading-logarithmic corrections from higher-loop contributions
must be included. These terms can be summed via renormalization
group techniques.  The result of these corrections is to reduce
the one-loop upper bound on $\mhl$. For $\ms\lsim {\cal O}(1~{ \rm
TeV})$, it is found that $\mhl\lsim 125$~GeV, where the maximum is
reached at large $\mha$ and $\tanb$ when $\ms$ is maximal and
$X_t\simeq \pm\sqrt{6}\ms$ (the so-called ``maximal-mixing'' value
for stop mixing). However, sub-leading two-loop corrections may
not be negligible, and a more complete two-loop computation is
required.

The full diagrammatic calculations lead to very complicated expressions
for the radiatively corrected value of $\mhl$.
Effective potential and effective field theory techniques
have been developed which can extract the dominant
contributions to the Higgs mass radiative corrections (when $\ms$ is
large), resulting in a simpler analytic expression for $\mhl$.
These methods also provide a natural setting for renormalization
group improvement.  Although the exact solution of the renormalization
group equations (RGEs) must be obtained numerically, the iterative
solution of the RGEs can easily yield simple analytic expressions for
the one-loop and two-loop leading logarithmic contributions to $\mhl$.
These leading logarithms can also
be obtained by expanding the complete
diagrammatic results in the limit of $\ms\gg\mt$, and this serves as an
important check of the various computations.

The effective potential method~\cite{quiros} provides an important tool
for evaluating the Higgs mass beyond the tree-level.
It can be used to provide a short-cut for
the calculation of certain combinations of Higgs boson two-point
functions that arise in the diagrammatic computation.
The effective field theory (EFT) approach~\cite{mhiggsOL1}
provides a powerful method for
isolating the leading terms of the Higgs mass radiative corrections when
$\ms\gg\mt$.  In this formalism, one matches the full supersymmetric
theory above $\ms$ to an effective Standard Model with supersymmetric
particles decoupled below $\ms$.  The Standard Model couplings
in the \msbar\ scheme are fixed
at $\ms$ by supersymmetric matching conditions.  Standard Model RGEs are
then used to evolve these couplings down to the electroweak
scale (either $\mt$ or $\mz$).
While the stops are decoupled at $\ms$, the stop mixing
(so-called ``threshold'') effects are incorporated by modifying the
matching conditions at $\ms$.  With the use of effective potential
techniques, the EFT formalism and the iteration of the RGEs to two-loops,
the leading contributions to the radiatively-corrected Higgs mass was
obtained in analytic form in
refs.~\cite{mhiggsrg1,mhiggsrg2,mhiggsrg3,mhiggsrg4}
These results included the
full one-loop leading logarithmic corrections and one-loop leading
squark-mixing threshold corrections, the two-loop leading
double-logarithmic
corrections and the two-loop leading logarithmic squark-mixing threshold
corrections up to ${\cal O}(h_t^2\alpha_s)$ and ${\cal O}(h_t^4)$, where
$h_t$ is the Higgs--top quark Yukawa coupling.

In order to extend the above results, genuine two-loop computations are
required.  The first two-loop diagrammatic computation was performed in
\Ref{mhiggseffpot1} in the limit of $\mha\gg\mz$ and $\tan\beta\gg 1$
(where $\mhl$ attains its maximal bound), where only terms of ${\cal
O}(h_t^4)$ and ${\cal O}(h_t^2\alpha_s)$ were evaluated, and all squark
mixing effects were neglected.
More recently, a more complete two-loop diagrammatic computation of the
dominant contributions at $\oaas$ to the neutral $\cp$-even Higgs boson
masses has been performed~\cite{mhiggsdiag1,mhiggsdiag2,mhiggsdiag3}.
This result was
obtained for arbitrary values of $\mha$, $\tanb$ and the stop-mixing
parameter $X_t$.  This two-loop
result, which had been obtained first in the on-shell scheme, was
subsequently combined~\cite{mhiggsdiag2,mhiggsdiag3}
with the complete
diagrammatic one-loop on-shell result of \Ref{mhiggsOL2} and the
leading two-loop Yukawa corrections of ${\cal O}(h_t^4)$ obtained
by the EFT approach~\cite{mhiggsrg1,mhiggsrg2,mhiggsrg3,mhiggsrg4}.
The resulting two-loop expression was then expressed in
terms of the top-quark mass in the \msbar\ scheme.
By comparing the final expression with the results obtained in
refs.~\cite{mhiggsrg1,mhiggsrg2,mhiggsrg3,mhiggsrg4},
it was shown that the upper bound on the lightest
Higgs mass was shifted upwards by up to 5~GeV, an effect that is
more pronounced in the low $\tan\beta$ region.

Besides the shift in the upper bound of $m_h$, apparent deviations
between the explicit diagrammatic two-loop calculation and the
results of the EFT computation were observed in the dependence
of $\mhl$ on the stop-mixing parameter $X_t$. While the value of
$X_t$ that maximizes the lightest $\cp$-even Higgs mass is
$(X_t)_{\rm max}\simeq \pm\sqrt{6}\,\ms \approx \pm 2.4\,\ms$ in the
results of refs.~\cite{mhiggsrg1,mhiggsrg2,mhiggsrg3,mhiggsrg4},
the corresponding on-shell two-loop diagrammatic computation found
a maximal value for $\mhl$ at $(X_t)_{\rm max}\approx
2\ms$.\footnote{A local maximum for $\mhl$ is also found for
$X_t\simeq -2\ms$, although the corresponding value of $\mhl$ at
$X_t\simeq +2\ms$ is significantly
larger~\cite{mhiggsdiag3,mhdiagcomp}.}
Moreover, in the
results of refs.~\cite{mhiggsrg1,mhiggsrg2,mhiggsrg3,mhiggsrg4},
$\mhl$ is symmetric under $X_t\to -X_t$ and has a (local) minimum
at $X_t=0$.  In contrast, the two-loop diagrammatic
computation yields $\mhl$
values for positive and negative $X_t$ that differ significantly from
each other and the local minimum in $\mhl$ is shifted slightly
away from $X_t= 0$~\cite{mhdiagcomp}.   A similar conclusion was
reached in \Ref{mhiggseffpot2}, which used an effective potential
calculation to extend the results of \Ref{mhiggseffpot1} to the case of
non-zero stop mixing.

A closer comparison of results of the EFT computation of the
radiatively-corrected Higgs mass and the two-loop diagrammatic
computation at first sight revealed a surprising discrepancy.
Namely, the two-loop leading logarithmic squark-mixing threshold
corrections at ${\cal O}(h_t^2\alpha_s)$ of the former do not
appear to match the results of the latter.  In this paper, we
shall show that this apparent discrepancy in the
leading-logarithmic contributions is caused by the different
renormalization schemes employed in the two approaches.  While the
original two-loop diagrammatic computations of \Ref{mhiggsdiag1}
were performed in an on-shell scheme, the results of the EFT
approach are most naturally carried out in the \msbar\ scheme. In
comparing results obtained within different renormalization
schemes in terms of the (not directly observable) parameters $X_t$
and $\ms$, one has to take into account the fact that these
parameters are renormalization-scheme dependent.  The effect of
this scheme dependence first enters into the calculation of $\mhl$
at the two-loop level. In order to allow a detailed comparison
between the results of the different approaches we derive
relations between these parameters in the two different schemes.
We apply these relations to re-express the diagrammatic on-shell
result in terms of \msbar\ parameters. In this way we show that
the leading logarithmic two-loop contributions in the two
approaches in fact coincide. The remaining numerical difference
between the diagrammatic calculation in the \msbar\ scheme and the
result obtained by the EFT approach can thus be identified with
new threshold effects due to non-logarithmic two-loop terms
contained in the diagrammatic result. We furthermore show that in
the analytic approximation employed in this paper, the dominant
numerical contribution of these terms can be absorbed into an
effective one-loop expression by choosing an appropriate scale for
the running top-quark mass in different terms of the expression.

This paper is organized as follows.  To simplify the analysis, we focus
completely on the radiatively corrected Higgs squared-mass in the
``leading $m_t^4$ approximation'' (in which only the dominant loop
corrections proportional to $m_t^2 h_t^2 \sim G_F m_t^4$ are kept).
In addition, we choose a very simple form for the stop squared-mass
matrix, which significantly simplifies the subsequent analysis while
maintaining the most important features of the general result.
In section~2 we sketch the derivation of the EFT result for the
radiatively-corrected Higgs mass of
refs.~\cite{mhiggsrg1,mhiggsrg2,mhiggsrg3,mhiggsrg4}
at ${\cal O}(m_t^2 h_t^2\alpha_s)$ in the limit of $\mha\gg\mz$.
The corresponding result of the two-loop diagrammatic computation, under
the same set of approximations, is outlined in section~3.
In order to compare the two results, we must convert on-shell quantities
to \msbar\ quantities.  In section~4
we derive the relations between the on-shell and the
\msbar\ values of the parameters $\mt$, $X_t$ and $\ms$ in the limit of
large $\ms$.  Details of the exact calculation are given in Appendix~A,
while explicit
relations between the on-shell and the \msbar\ parameters up to
${\cal O}(m_t^4/M_S^4)$ are given in Appendix~B.
In section~5 the diagrammatic
on-shell result is expressed in terms of \msbar\ parameters and compared
to the result of the EFT computation of section~2. The logarithmic
contributions are
shown to coincide, and the remaining difference caused by
non-logarithmic two-loop terms is analyzed.
We argue that the remaining difference can be minimized by improving the
EFT computation by taking into account the stop-mixing threshold
contribution to the running top-quark mass.   In addition,
we demonstrate that in a simple analytic approximation to the Higgs
mass, one can absorb the dominant two-loop contributions into an
effective one-loop expression.
In section~6, we summarize our results and discuss suggestions for
future improvements.

\section{Effective Field Theory Approach}

At the tree level, the mass matrix of the neutral $\cp$-even Higgs
bosons in the basis of weak eigenstates of definite hypercharge $-1$
and $+1$ respectively can be expressed
in terms of $\mz$, $\mha$ and $\tan\beta\equiv v_2/v_1$ as follows:
\be
{\cal M}_H^{2, {\rm tree}} = \left( \begin{array}{cc}
     \mha^2 \sin^2\beta + \mz^2 \cos^2\beta &
     -(\mha^2 + \mz^2) \sin\beta \cos\beta \\
     -(\mha^2 + \mz^2) \sin\beta \cos\beta &
     \mha^2 \cos^2\beta + \mz^2 \sin^2\beta \end{array} \right) .
\label{eq:mhmatrix}
\ee
Diagonalizing this mass matrix yields the tree-level prediction for the
lightest neutral $\cp$-even Higgs-boson mass
\be
\label{mh2tree}
\mhl^{2,{\rm tree}} = \half \left[ \mha^2 + \mz^2
         - \sqrt{(\mha^2 + \mz^2)^2 - 4 \mz^2 \mha^2 \cos^2 2\beta}\,
\right]\,.
\ee

For simplicity, we shall consider the limit of $\mha\gg\mz$ and large
supersymmetry-breaking masses characterized by a scale $\ms$.  Then,
at energy scales below $\ms$, the effective low-energy theory consists
of the Standard Model with one Higgs doublet.  The corresponding Higgs
squared-mass at tree-level is given by
$\mhl^{2,{\rm tree}}= \mz^2 \cos^2 2 \beta$.
The dominant contributions to the radiatively-corrected
Higgs mass within the EFT approach is based on the evaluation
of the effective quartic Higgs self-coupling, $\lambda$,
evaluated at the scale $Q = \mtbar$.
The value of $\lambda(\ms)$ is fixed by the supersymmetric boundary
condition, although this value is slightly modified by
one-loop threshold effects (denoted below by $\Delta_{\rm th}\lambda$),
due to the decoupling of squarks at $\ms$ with non-zero mixing.
One then employs the Standard Model RGEs to
obtain $\lambda(Q)$.   Finally, the Higgs mass is obtained via
$\mhl^2=2\lambda(\mtbar) v^2(\mtbar)$ [where $v=174$~GeV is the Higgs
vacuum expectation value].
The mass $\mtbar$ denotes the running top-quark mass in the
\msbar\ scheme at the scale $\mt$. It is related to the
on-shell (or pole) top-quark mass
$M_t \equiv m_t^{\OS}$ by the following relation
\be
\label{eq:mtmsbar}
\mtbar \equiv \mtsmreg(\mt) = \frac{M_t}{1+\frac{4}{3\pi}\alpha_s(M_t)},
\ee
where we have only included
the QCD corrections to leading order in $\alpha_s$. In \eq{eq:mtmsbar},
the subscript `SM' indicates that the running mass is defined
in the usual way, {\it i.e.} in terms of the pure Standard Model
(gluonic) contributions in the (modified) minimally
subtracted dimensional regularization (DREG) scheme~\cite{dreg}.

The most important contributions to the mass of the lightest $\cp$-even
Higgs boson arise from the $t$--$\Stop$ sector of the MSSM, which
is characterized by the following squared-mass matrix
\be
\label{eq:stopmatrix}
{\cal M}^2_{\stop} =
  \left( \begin{array}{cc}
   \MstL^2 + m_t^2 + \cos 2\beta (\frac{1}{2} - \frac{2}{3} s_W^2) \mz^2 &
      \mt X_t \\
      \mt X_t &
      \MstR^2 + \mt^2 + \frac{2}{3} \cos 2\beta s_W^2 \mz^2
  \end{array} \right) \,,
\ee
where $X_t\equiv A_t - \mu\cot\beta$.
The $\Stop$-masses $\mste$, $\mstz$
and the mixing angle $\tst$ are determined at tree-level by
diagonalizing ${\cal M}^2_{\stop}$.
Neglecting  the numerically small contributions proportional to $\mz^2$
in the stop squared-mass matrix and setting
\be
\MstL = \MstR \equiv \msq,\;\; M_S^2 \equiv\msq^2 + m_t^2
\ee
leads to the simplified mass matrix
\be
\label{stopmassmatrix}
{\cal M}^2_{\stop} =
  \left( \begin{array}{cc}
    \ms^2  &  \mt X_t \\
    \mt X_t & \ms^2
\end{array} \right) .
\ee
In this approximation, the
$\Stop$-masses and the mixing angle are given by
\bea
\label{mstopins}
\mste^2 &=& \ms^2 - |\mt X_t| , \nonumber \\
\mstz^2 &=& \ms^2 + |\mt X_t| , \\
\label{mixangins}
\tst &=& \left\{ \renewcommand{\arraystretch}{1.3}
         \begin{array}{r@{~~~}l}
               \frac{\pi}{4} & \mbox{for } X_t < 0 \\
               %0 & \mbox{for } X_t = 0\,, \\
               -\frac{\pi}{4} & \mbox{for } X_t > 0\,,
         \end{array} \right.
         \renewcommand{\arraystretch}{1}
\eea
where by definition, $\mste\leq\mstz$.

The one-loop threshold corrections to the quartic Higgs self-coupling,
induced by the decoupling of stops,
lead to  a change of the effective quartic Higgs self-coupling at
the scale $\ms$,
\be
\label{inicial}
\lambda(\ms)=
\quarter(g^2+g'^2) \cos^2 2 \beta
+\Delta_{\rm th}\lambda ,
\ee
where the first term is the tree-level value of the
quartic Higgs self-coupling in the effective low-energy Standard
Model and
the second term is the effect of the one-loop threshold corrections
at the scale $\ms$~\cite{mhiggsOL1,mhiggsrg3},
\be
\label{expumbral}
\Delta_{\rm th}\lambda=
\frac{3}{8\pi^2}
h_t^2\left\{\left[h_t^2-\nicefrac{1}{8}(g^2+g'^2)\right]
\left(\frac{X_t^2}{\ms^2}\right)
-\nicefrac{1}{12}h_t^2\left(\frac{X_t^4}{\ms^4}\right)\right\}
+\cdots \,,
\ee
where all couplings in \eq{expumbral} should be
evaluated at the scale $\ms$.  The running Higgs--top quark Yukawa
coupling is related to the \msbar\ running top quark mass:
\be  \label{htdef}
\mtbar(\mu)=h_t(\mu) v(\mu)\,,
\ee
where the running Higgs vacuum expectation value,
$v^2(\ms) = v^2(\mbart)\,\xi^{-2}(\mbart)$,
is governed by the Higgs field anomalous dimension
\be
\label{anomdim}
\xi(\mbart)=1+\frac{3}{32\pi^2}h_t^2(\mbart)
\ln\left(\frac{\ms^2}{\mbart^2}\right)\,.
\ee
In obtaining \eq{expumbral}, an expansion in the variable
\be
\label{eq:dst}
\dst\equiv \frac{|\mt X_t|}{\ms^2} =
       \frac{\mstz^2 - \mste^2}{\mstz^2 + \mste^2}, \quad
0 \le \dst < 1,
\ee
has been performed. Terms not explicitly exhibited
in \eq{expumbral} denote the contributions from
higher powers in $m_t/\ms$ and $X_t^2/\ms^2$, which
arise from the contributions of the
$t$--$\Stop$ sector.
Contributions from other supersymmetric-breaking
sectors have been omitted for simplicity of the presentation.
These contributions typically contribute no more than a few GeV to the
radiatively-corrected Higgs mass.

As it was shown in \Refs{mhiggsrg3}{mhiggsrg4},
one can obtain the two-loop leading-logarithmic
correction by expanding the parameter $\lambda$
up to order $[\ln(\ms^2/\mbart^2)]^2$,
\bea
\label{efflamix2}
\leff(\mbart)& = & \leff(\ms)-\beta_{\leff}(\ms)t
+\half\beta'_{\leff}(\mbart)t^2 +\cdots \nonumber \\
& = & \leff(\ms)-\beta_{\leff}(\mbart)t
-\half\beta'_{\leff}(\mbart)t^2+\cdots
\eea
where $\leff(\ms)$ is given by \eq{inicial}
and
\be
\label{escala}
t\equiv \ln\frac{\ms^2}{\mbart^2}\,.
\ee
Following \Ref{mhiggsrg3}, we define
$\beta_{\leff} = a_{\leff} \leff + b_{\leff}$. Therefore
\be \label{lambdarun}
\lambda(\mbart) = \lambda(\ms)  \left[1
- a_{\leff}(\mbart) \; t \right] - b_{\leff}(\mbart) \; t \;
\left[1
- a_{\leff}(\mbart) \; t \right] -\half\beta'_{\leff}(\mbart)t^2\,.
\ee
Here, $1 - a_{\leff}(\mbart) \;t = \xi^{-4}(\mbart)$, where
$\xi$ is  the Higgs field anomalous dimension [\eq{anomdim}].
Multiplying \eq{lambdarun} by $2v^2(\mtbar)$, we obtain an equation for
the Higgs squared-mass in the low-energy theory, which takes the
following form:
\be
\label{decmasa}
m_h^2(\mbart) = m_h^2(\ms)\;\xi^{-2}(\mbart)+
\Delta_{\rm rad\; }m_h^2(\mbart)\,,
\ee
which defines the quantity $\Delta_{\rm rad\; }m_h^2(\mbart)$.  In
\eq{decmasa},
\be
\label{masafroz}
\mhl^2(\ms) = 2 \lambda(\ms) v^2(\ms)\,,
\ee
where $\lambda(\ms)$ is given in \eq{inicial}
with all couplings and masses evaluated at the scale~$\ms$.

In the present analysis, we are working in the approximation of
$h_b=g=g'=0$.  That is, we focus
only on the Higgs--top quark Yukawa and QCD coupling effects.
The relevant $\beta$-functions for
$\lambda$, $g_3^2$ and $h_t^2$ at scales below the scale $\ms$
are given by~\cite{smrges}
\bea
16\pi^2\beta_{\lambda} &\equiv & 6(\lambda^2 +\lambda h_t^2 - h_t^4)
+\frac{h_t^4}{8 \pi^2}\left(15 h_t^2-16g_3^2\right) \,,\\
16 \pi^2 \beta_{h_t^2} &\equiv & h_t^2 \left(\nicefrac{9}{2} h_t^2- 8g_3^2
\right)\,,\\
16 \pi^2 \beta_{g_3^2} &\equiv &\left( -11 + \nicefrac{2}{3} N_f\right)
g_3^4 \,,
\eea
where $\beta_{X}\equiv dX/d\ln Q^2$ and $N_f$
is the number of quark flavors with masses less than~$Q$
({\it e.g.}, $N_f =6$ for scales between $m_t$ and $\ms$).
Observe that we have included the dominant strong gauge coupling
two-loop contribution to the $\beta$ function of the quartic Higgs
self-coupling,
since it will contribute once we include all two-loop
leading-logarithmic corrections.

Using the above expressions, it is simple to find an approximate
formula for the lightest $\cp$-even Higgs mass in the large $\mha$
limit.  First, one obtains
\bea
\label{radmas}
\Delta_{\rm rad\; }m_h^2(\mbart)& = &
\frac{3}{4\pi^2}\frac{\mbart^4}{v^2(\mbart)}\;t
\left[1+\frac{1}{16\pi^2}\left(\nicefrac{3}{2}h_t^2-32\pi\alpha_s\right)t
\right]\,,
\eea
where all couplings in \eq{radmas} are evaluated at
the scale $Q^2=\mbart^2$.  To complete the computation of
$m_h^2(\mbart)$ [\eq{decmasa}], one must evaluate $\mhl^2(\ms)$
[see \eq{masafroz}] in terms of low-energy parameters. This is
accomplished by using one-loop renormalization group evolution to relate
$\lambda(\ms)v^2(\ms)$ to $\lambda(\mtbar)v^2(\mt)$. In this way,
one finally arrives at the expression\footnote{In the
``leading $m_t^4$ approximation'' that is employed here, there is no
distinction between $\mhl(\mtbar)$ and the on-shell (or
pole) Higgs mass, $\mhl$.}
\bea
\label{mhsm}
m_h^2& = &
  \mhl^{2,{\rm tree}}
  +   \frac{3}{4\pi^2}\frac{\mbart^4}{v^2}\left\{t+ \frac{X_t^2}{\ms^2}
                  \left(1 - \frac{X_t^2}{12 \ms^2} \right)
%\frac{1}{2}\tilde{X}_t + t
\right. \nonumber \\
&&\!\!\!\!\!\! {} + \left.
\frac{1}{16\pi^2}\left(\frac{3}{2}\frac{\mbart^2}{v^2}-32\pi\alpha_s
\right)\left[\frac{2 X_t^2}{\ms^2}
                  \left(1 - \frac{X_t^2}{12 \ms^2} \right)t+
t^2\right]  + \left(\frac{ 4 \alpha_s}{3 \pi}
  - \frac{5h_t^2}{16 \pi^2}\right) t \right\} \,.
\eea
The last two terms in \eq{mhsm}
reflect the two-loop single logarithmic dependence induced
by the two-loop
$\beta$-function contribution to the running of
the quartic Higgs self-coupling.  It is interesting to note
that these two terms are numerically close in size, and
they tend to cancel each other in the computation of the Higgs mass.
\Eq{mhsm} differs from the one presented in \Ref{mhiggsrg2}
only in the inclusion of these terms, which although
sub-dominant compared to the remaining terms, should be
kept for comparison with the diagrammatic result.

The full two-loop corrections to $\mhl^2$ at ${\cal O}(m_t^2 h_t^4)$
have not yet been calculated in the diagrammatic approach; thus we
neglect terms of this order in what follows.\footnote{As noted
below \eq{mhsm}, terms of ${\cal O}(m_t^2 h_t^4)$ can be as numerically
important as terms of ${\cal O}(m_t^2 h_t^2\alpha_s)$.  Hence, in a
complete phenomenological analysis, one should not neglect terms of the
former type.}
%\mua
With a slight rewriting of \eq{mhsm}
we finally obtain the expression that will be compared with the
diagrammatic result in the following sections:
\bea
\lefteqn{
\mhl^2 = \mhl^{2,{\rm tree}} +
  \frac{3}{2} \frac{G_F \sqrt{2}}{\pi^2} \mtbar^4 \left\{
  - \ln\left(\frac{\mtbar^2}{\Msbarii} \right) +
    \frac{\Xtbarii}{\Msbarii}
  \left(1 - \frac{1}{12} \frac{\Xtbarii}{\Msbarii} \right)
  \right\} \non } \\
&& {} - 3 \frac{G_F \sqrt{2}}{\pi^2} \frac{\alpha_s}{\pi} \mtbar^4
  \left \{\ln^2 \left(\frac{\mtbar^2}{\Msbarii} \right) +
    \left[\frac{2}{3} - 2 \frac{\Xtbarii}{\Msbarii}
          \left(1 - \frac{1}{12} \frac{\Xtbarii}{\Msbarii} \right) \right]
    \ln\left(\frac{\mtbar^2}{\Msbarii} \right) \right \},
\label{eq:mhrg}
\eea
where we have introduced the notation $\Msbar, \Xtbar$ to emphasize that
the corresponding quantities are \msbar\ parameters, which are
evaluated at the scale $\mu = \ms$:
\be
\label{eq:MsXtbar}
\Msbar \equiv \ms^{\MS}(\ms), \quad
\Xtbar \equiv X_t^{\MS}(\ms),
\ee
and
$\mtbar \equiv \mtsmreg(\mt)$ as defined in \eq{eq:mtmsbar}.

%%%%%%%%%%%%%%%%%%%%%%%%%%%%%%%%%%%%%%%%%%%%%%%%%%%%%%%%%%%%%%%%%%%%%%%%%%%%%

\section{Diagrammatic calculation}
\label{sec:diagramm}

In the diagrammatic approach the masses of the $\cp$-even
Higgs bosons are obtained by evaluating loop corrections to the
$h$, $H$ and $hH$-mixing propagators. The masses of the two
$\cp$-even Higgs bosons, $\mhl$ and $\mhh$, are
determined as the poles of this propagator matrix, which are given by the
solution of
\be
\label{eq:mhpole}
\left[q^2 - \mhl^{2,{\rm tree}} + \hat\Sigma_{hh}(q^2) \right]
\left[q^2 - \mhh^{2,{\rm tree}} + \hat\Sigma_{HH}(q^2) \right] -
\left[\hat\Sigma_{hH}(q^2)\right]^2 = 0 ,
\ee
where $\hat\Sigma_{hh}(q^2)$, $\hat\Sigma_{HH}(q^2)$,
$\hat\Sigma_{hH}(q^2)$ denote the renormalized Higgs boson
self-energies. In \Ref{mhiggsdiag1} the dominant two-loop
contributions to the masses of the $\cp$-even Higgs bosons of $\oaas$
have been evaluated. These corrections, obtained in the
on-shell scheme, have been combined in
\Refs{mhiggsdiag2}{mhiggsdiag3} with the complete one-loop
on-shell result of \Ref{mhiggsOL2} and the two-loop
corrections of ${\cal O}(m_t^2 h_t^4)$ given in
refs.~\cite{mhiggsrg2,mhiggsrg3,mhiggsrg4}.

The diagrammatic two-loop calculation of \Ref{mhiggsdiag1} involves a
renormalization in the Higgs sector up to the two-loop level and a
renormalization in the stop sector up to $\oas$.
In the on-shell scheme, the renormalization in the stop sector
is performed such that the
$\Stop$-masses $\mste$, $\mstz$ correspond to the poles of the
propagators, {\it i.e.}
\be
\mathrm{Re} \, \hat\Sigma_{\stop_1\stop_1}(\mste^2) = 0, \quad
\mathrm{Re} \, \hat\Sigma_{\stop_2\stop_2}(\mstz^2) = 0
\ee
for the renormalized self-energies. In Ref.~\cite{mhiggsdiag1}
the renormalization condition
\be  \label{stop12}
\mathrm{Re} \, \hat\Sigma_{\stop_1\stop_2}(\mste^2) = 0
\ee
has been chosen to define the stop mixing angle.\footnote{In this
paper, our analysis is presented in a simplified model of
stop mixing, where the tree-level stop squared-mass
matrix given by \eq{stopmassmatrix}.  In this case, $\widetilde t_1$ and
$\widetilde t_2$ are states of definite parity at tree-level.   Since
parity is preserved to all orders in $\alpha_s$, it follows that
$\hat\Sigma_{\stop_1\stop_2}(p^2)=0$ (when electroweak corrections are
neglected) and \eq{stop12} is trivially satisfied.}
% As will become clear below, the precise form of
%this renormalization condition is unimportant for the comparison of the
%on-shell with the \msbar\ results.
In \Ref{mhdiagcomp} a compact analytic approximation has been
derived from the rather complicated diagrammatic two-loop result by
performing an expansion in $\Delta_{\tilde t}$ [\eq{eq:dst}]
of the $t$--$\Stop$ sector contributions.

The diagrammatic two-loop corrections to the Higgs mass also depend
non-trivially on the gluino mass, which is a free input parameter of the
supersymmetric model.  In the EFT approach described in section 2,
the gluino is decoupled at the same scale as the stops.  Thus, in order
to compare the results of the EFT and diagrammatic approaches, one must
take $\mgl\simeq{\cal O}(\ms)$.  In this paper, we have chosen
\be
\label{eq:mgl}
\mgl = \msq = \sqrt{\ms^2 - \mt^2}\,.
\ee
For the one-loop contributions from the other sectors of the MSSM the
leading logarithmic approximation has been
used~\cite{mhiggsOL1,mhiggsrg4}. In this approximation,
the momentum dependence in \eq{eq:mhpole} is neglected
everywhere. The resulting expression can thus be written as a correction
to the tree-level mass matrix [\eq{eq:mhmatrix}]. The expression
for $\mhl^2$ in this
approximation is obtained by diagonalizing the loop-corrected mass
matrix. The compact analytic expression derived in this way, which is
valid for arbitrary values of $\mha$, has been shown to approximate
the full diagrammatic result for $\mhl$ rather well, typically within
about 2~GeV for most parts of the MSSM parameter
space~\cite{mhdiagcomp}.

In the following we will restrict ourselves to the contribution of the
$t$--$\widetilde t$ sector. In order to perform a simple comparison with
the EFT approach of section 2,
we only consider the dominant one-loop and two-loop
terms of ${\cal O}(\mt^2 h_t^2)$ and ${\cal O}(m_t^2 h_t^2 \alpha_s)$,
respectively. We focus on the case $\mha \gg \mz$, for which
the result for $\mhl^2$ can be expressed in a particularly compact form,
\be
\label{eq:mhdiagos}
\mhl^2 = \mhl^{2,{\rm tree}} + \mhl^{2,{\alpha}} +
\mhl^{2,{\alpha\alpha_s}},
\ee
and neglect the non-leading terms of ${\cal O}(\mz^2/\mha^2)$.
Moreover, assuming that $\ms \gg M_t$ and neglecting the
non-leading
terms of ${\cal O}(M_t/\ms)$ and ${\cal O}(\mz^2/M_t^2)$, one obtains
the following simple result for the one-loop and two-loop contributions
\bea
\mhl^{2,{\alpha}} &=&
  \frac{3}{2} \frac{G_F \sqrt{2}}{\pi^2} M_t^4 \left\{
  - \ln\left(\frac{M_t^2}{\ms^2} \right)
  + \frac{X_t^2}{\ms^2}
  \left(1 - \frac{1}{12} \frac{X_t^2}{\ms^2} \right)
  \right\}, % + {\cal O}\left(\frac{M_t}{\ms}\right),
\label{eq:mh1ldiagos} \\
\mhl^{2,{\alpha\alpha_s}} &=&
  - 3 \frac{G_F \sqrt{2}}{\pi^2} \frac{\alpha_s}{\pi} M_t^4
  \left\{\ln^2 \left(\frac{M_t^2}{\ms^2} \right)
  - \left(2 + \frac{X_t^2}{\ms^2} \right)
  \ln\left(\frac{M_t^2}{\ms^2} \right)
- \frac{X_t}{\ms} \left(2 - \frac{1}{4} \frac{X_t^3}{\ms^3} \right)
  \right\} . \non \\
&& \label{eq:mh2ldiagos}
\eea
The corresponding formulae, in which terms up to ${\cal O}(M_t^4/\ms^4)$
are kept, can be found in Appendix~B [see
\eqns{eq:mh1ldiagosApp}{eq:mh2ldiagosApp}].

In \eqns{eq:mh1ldiagos}{eq:mh2ldiagos} the parameters
$M_t$, $\ms$, $X_t$ are on-shell quantities.
Using \eq{eq:mtmsbar}, the on-shell result for $\mhl^2$
[\eqs{eq:mhdiagos}{eq:mh2ldiagos}] can easily be rewritten
in terms of the running top-quark mass $\mtbar$.
While this reparameterization does not change the
form of the one-loop result, it induces an extra contribution at
$\oaas$. Keeping %for simplicity only terms that are not
again only terms that are not
suppressed by powers of $\mtbar/\ms$, %${\cal O}(\mtbar^0/\ms^0)$,
the resulting expressions read
\bea
\mhl^{2,{\alpha}} &=&
  \frac{3}{2} \frac{G_F \sqrt{2}}{\pi^2} \mtbar^4 \Biggl\{
  - \ln\left(\frac{\mtbar^2}{\ms^2} \right) +
  \frac{X_t^2}{\ms^2}
  \left(1 - \frac{1}{12} \frac{X_t^2}{\ms^2} \right)
  \Biggr\} , % + {\cal O}\left(\frac{\mtbar}{\ms}\right),
  \label{eq:mh1ldiagosmtbar} \\
\mhl^{2,{\alpha\alpha_s}} &=&
  - 3 \frac{G_F \sqrt{2}}{\pi^2} \frac{\alpha_s}{\pi} \mtbar^4
  \Biggl\{\ln^2 \left(\frac{\mtbar^2}{\ms^2} \right) +
  \left(\frac{2}{3} - \frac{X_t^2}{\ms^2} \right)
  \ln \left(\frac{\mtbar^2}{\ms^2} \right)\non  \\
&& {} + \frac{4}{3} - 2 \frac{X_t}{\ms} - \frac{8}{3} \frac{X_t^2}{\ms^2}
   + \frac{17}{36} \frac{X_t^4}{\ms^4} \Biggr\} ,
   %+ {\cal O}\left(\frac{\mtbar}{\ms}\right),
\label{eq:mh2ldiagosmtbar}
\eea
in accordance with the formulae given in \Ref{mhdiagcomp}.

We now compare the diagrammatic result expressed in terms of the parameters
$\mtbar$, $\ms$, $X_t$
[\eqns{eq:mh1ldiagosmtbar}{eq:mh2ldiagosmtbar}]
with the EFT result [\eq{eq:mhrg}] which is given
in terms of the \msbar\ parameters $\mtbar$, $\Msbar$, $\Xtbar$
[\eq{eq:MsXtbar}].  While the $X_t$--independent logarithmic terms
are the same in both the diagrammatic
and EFT results, the corresponding logarithmic terms at two-loops that
are proportional to powers of $X_t$ and $\Xtbar$, respectively,
are different.  Furthermore, \eq{eq:mh2ldiagosmtbar} does not
contain a logarithmic term proportional to $X_t^4$,
%there exists no logarithmic term proportional to $X_t^4$ in
%Eq.~(\ref{eq:mh2ldiagosmtbar}),
while the corresponding  term proportional to $\Xtbariv$ appears in
\eq{eq:mhrg}.
To check whether these results are consistent, one must relate the
on-shell and \msbar\ definitions of the parameters $\ms$ and $X_t$.

Finally, we note that the non-logarithmic terms contained in
\eq{eq:mh2ldiagosmtbar}
correspond to genuine two-loop contributions
that are not present in the EFT result of \eq{eq:mhrg}.
They can be interpreted as a two-loop
finite threshold correction to the quartic Higgs self-coupling in the
EFT approach.  In particular, note that
\eq{eq:mh2ldiagosmtbar} contains a term that is
linear in $X_t$
This is the main source of the asymmetry in the two-loop corrected
Higgs mass under $X_t\to -X_t$ obtained by the diagrammatic method.
The non-logarithmic terms in \eq{eq:mh2ldiagosmtbar} give rise to
a numerically significant increase of the
maximal value of $\mhl$ of about 5~GeV in this approximation.

%%%%%%%%%%%%%%%%%%%%%%%%%%%%%%%%%%%%%%%%%%%%%%%%%%%%%%%%%%%%%%%%%%%%%%%

\section{\boldmath{On-shell and $\overline{\rm MS}$ definitions
of $\ms$ and $X_t$}}

Since the parameters $p = \{\mste^2, \mstz^2, \tst, \mt\}$ of the $t$--$\Stop$
sector are renormalized differently in different schemes,
the parameters $\ms$ and $X_t$ also have a different meaning in these
schemes. In order to derive the relation between these parameters in the
\msbar\ and in the \os\ scheme we start from the observation that at
lowest order the parameters $p$ are the same in both schemes, {\it i.e.}
$p = p^{\OS} = p^{\MS}$ in lowest order. Expressing the bare parameters
in terms of the renormalized parameters and the counterterms leads to
\be
p^{\MS} + \de p^{\MS} = p^{\OS} + \de p^{\OS} .
\ee
Here $\de p^{\OS}$ is the \os\ counterterm in $D\equiv 4 - 2\epsilon$
dimensions, and according to the \msbar\ prescription $\de p^{\MS}$
is given just by the pole part of $\de p^{\OS}$, {\it i.e.} the
contribution
proportional to $1/\epsilon - \gamma_{\mathrm E} + \ln 4 \pi$, where
$\gamma_{\mathrm E}$ is Euler's constant. The \msbar\
parameters are thus related to the \os\ parameters by%in the following way
\be
p^{\MS} = p^{\OS} + \De p,
\label{eq:paramrel}
\ee
where $\De p \equiv \de p^{\OS} - \de p^{\MS}$ is finite in the limit
$D \to 4$ and contains the \msbar\ scale $\mu$ which can be chosen
appropriately.  In this paper, we only need to know $\Delta p$ to
${\cal O}(\alpha_s)$ one-loop accuracy.
In the following we will compare the result for $\mhl$
expressed in terms of the on-shell parameters $p^{\OS}$ with results
for $\mhl$ in terms of the corresponding \msbar\ parameters $p^{\MS}$,
which are related to $p^{\OS}$ as in \eq{eq:paramrel}.

In the EFT approach, the parameters $\ms$ and $X_t$ are running
parameters evaluated at the scale $\mu = \ms$ [\eq{eq:MsXtbar}].
In the simplified model for the stop squared-mass matrix given by
\eq{stopmassmatrix},
the relations between the parameters $X_t$ and $\ms$ in the on-shell and
\msbar\ scheme are obtained using
\bea
\label{eq:mstrels1}
\mste^{2, \OS} = \ms^{2, \OS} \mp M_t X_t^{\OS} \,, && \quad
      \mstz^{2, \OS} = \ms^{2, \OS} \pm M_t X_t^{\OS} \,, \\
\mste^{2, \MS} = \Msbarii \mp \mtbar(\ms)\Xtbar \,, &&
\quad
      \mstz^{2, \MS} = \Msbarii \pm \mtbar(\ms) \Xtbar \,,
\label{eq:mstrels2}
\eea
where we have written $\mtbar(\ms)\equiv
\mt^{\MS}(\ms)$ and $M_t\equiv \mt^{\OS}$ as in section 2.  In
both \eqns{eq:mstrels1}{eq:mstrels2}, the upper and lower signs
refer to $X_t^{\OS} > 0$ and $X_t^{\OS} < 0$, respectively. In the
model of stop mixing under consideration, there is no shift in the
scalar top mixing angle to all orders in $\alpha_s$, from which it
follows that $|\tst^{\OS}| = |\tst^{\MS}|$.

Inserting the relation between $\mste^2, \mstz^2$ in the on-shell and
the \msbar\ scheme into \eqns{eq:mstrels1}{eq:mstrels2}
yields up to first order in $\alpha_s$
\bea
\label{eq:msrel}
\Msbarii &=& \ms^{2, \OS} +
     \half \left( \De \mste^2 + \De \mstz^2 \right) , \\
\Xtbar &=& X_t^{\OS} \frac{M_t}{\mtbar(\ms)}
\pm \frac{1}{2 \mt} \left( \De \mstz^2 - \De \mste^2 \right)\,,
\label{eq:xtrel}
\eea
where again the upper and lower sign in the last equation refers to
$X_t^{\OS} > 0$ and $X_t^{\OS} < 0$, respectively. In the second term of
\eq{eq:xtrel} it is not necessary to distinguish (at
one-loop) between
$\mtbar(\ms)$ and $M_t$, since $\De \mste^2$, $\De \mstz^2$ are
$\oas$ quantities; hence, the generic symbol $\mt$ is used here.

In Appendix~A, we have obtained explicit results for $\De
\mste^2$, $\De \mstz^2$ and $M_t/\mtbar(\ms)$. Inserting the
appropriate expressions for these quantities into \eq{eq:xtrel}, one
observes that the functional form for $\Xtbar$ is the same for
$X_t^{\OS} > 0$ and $X_t^{\OS} < 0$ [{\it i.e.}, the sign
difference in \eq{eq:xtrel} is compensated by the term $(\De
\mstz^2 - \De \mste^2$)].  As a result, it is no longer necessary
to distinguish between these two cases. The case $X_t^{\OS} = 0$
(which formally would have to be treated separately) is understood
as being included in \eq{eq:xtrel}.

Using the expansions given in Appendix~A and setting the gluino mass
according to \eq{eq:mgl}, we obtain to leading order in $\mt/\ms$
\bea
\Msbarii &=& \ms^{2, \OS}
 - \frac{8}{3} \frac{\alpha_s}{\pi} \ms^2 \,,
\label{eq:msms} \\
\Xtbar &=& X_t^{\OS} \frac{M_t}{\mtbar(\ms)} +
  \frac{8}{3} \frac{\alpha_s}{\pi} \ms \,.
\label{eq:xtms}
\eea
As previously noted, it is not necessary to specify the definition of
the parameters that appear in the ${\cal O}(\alpha_s)$ terms.  Thus, we
use the generic symbol $\ms^2$ in the ${\cal O}(\alpha_s)$ terms of
\eqs{eq:msms}{eq:xtms}.  The corresponding results including terms
up to ${\cal O}\left(\mt^4/\ms^4\right)$ can be found in Appendix~B.

Finally, we need to evaluate the ratio $M_t/\mtbar(\ms)$.
The relevant expression is given in \eq{mtmsos}.  Using the expansions
given at the end of Appendix~A, we find to leading order in $\mt/\ms$
\be
\mtbar(\ms) =
  \mtbar \left[1 +
\frac{\alpha_s}{\pi}
    \ln\left(\frac{\mt^2}{\ms^2}\right)
   +  \frac{\alpha_s}{3\pi} \frac{X_t}{\ms} \right]\,,
\label{eq:mtsusy}
\ee
where $\mtbar\equiv \mtsmreg(\mt)$ is given in terms of $M_t$ by
\eq{eq:mtmsbar}.
The corresponding formula, where terms up to
${\cal O}\left(\mt^4/\ms^4\right)$ are kept, can be found at the end of
Appendix~B.
Note that the term in \eq{eq:mtsusy} that is proportional to $X_t$ is a
threshold correction due to the supersymmetry-breaking stop-mixing
effect. Inserting the result of \eq{eq:mtsusy} into \eq{eq:xtms} yields:
\be \label{eq:xtrelsusylead}
\Xtbar  = X_t^{\OS} + \frac{\alpha_s}{3 \pi} \ms
   \left[8 + \frac{4X_t}{\ms} - \frac{X_t^2}{\ms^2} - \frac{3X_t}{\ms}
\ln\left(\frac{\mt^2}{\ms^2}\right)
   \right]\,.
\ee
It is interesting to note that $\Xtbar\neq 0$ when $X_t^{\OS}= 0$.
Moreover, it is clear from \eq{eq:xtrelsusylead} that the relation
between $X_t$ defined in the on-shell and the \msbar\ schemes
includes a leading logarithmic effect, which has to be taken into
account in a comparison of the leading logarithmic contributions
in the EFT and the two-loop diagrammatic results.
%Eq.~(\ref{eq:xtrelsusylead}) contains an extra non-logarithmic term
%proportional to $x_t^2$
%compared to Eq.~(\ref{eq:xtrelreglead}). In the viewpoint of the RG
%  calculation, it is induced by the
%contributions in Eq.~(\ref{eq:mtsusy}) which arise from the decoupling
%of the $\Stop$ states.

The above results are relevant for calculations in the full theory in
which the effects of the supersymmetric particles are fully taken into
account.  However, in effective field theory below $\ms$, one must
decouple the supersymmetric particles from the loops and compute with
the Standard Model spectrum.  Thus, it will be useful to define a
running \msbar\ top-quark mass in the effective Standard Model,
$\mtsmreg(\mu)$, which to ${\cal O}(\alpha_s)$ is given by:
\be
\mtsmreg(\mu) =
  \mtbar \left[1 + \frac{\alpha_s}{\pi}
    \ln\left(\frac{\mt^2}{\mu^2}\right) \right] \,.
     \label{eq:mtsmreg}
\ee
At the scale $\ms$, we must match this result onto the expression for
$\mtbar(\ms)$ as computed in the full theory [\eq{eq:mtsusy}].  The
matching is discontinuous at $\mu=\ms$ due to the threshold corrections
arising from stop mixing effects.

In comparing with the EFT results of
refs.~\cite{mhiggsrg1,mhiggsrg2,mhiggsrg3,mhiggsrg4},
it should be noted that the threshold correction in \eq{eq:mtsusy}
were omitted.  This is relevant, since
$\mtbar(\ms)$ [or the related quantity $h_t(\ms)$, see \eq{htdef}]
appears in the threshold correction to the quartic Higgs self-coupling
$\lambda(\ms)$ [\eqns{inicial}{expumbral}].  In
refs.~\cite{mhiggsrg1,mhiggsrg2,mhiggsrg3,mhiggsrg4},
$\mtbar(\ms)$ is re-expressed in terms of $\mtbar(\mt)$ by using
\eq{eq:mtsmreg} rather than \eq{eq:mtsusy}.  As a result, a two-loop
non-logarithmic term proportional to $X_t$ is missed in the computation
of $\mhl$.  Such a term is of the same order as the
two-loop threshold correction to the quartic Higgs self-coupling, which
were also neglected in refs.~\cite{mhiggsrg1,mhiggsrg2,mhiggsrg3,mhiggsrg4}.
However, in this work we do not neglect the latter.  Hence, it would be
incorrect to use \eq{eq:mtsmreg} in the evaluation of $\mtbar(\ms)$.
In section~5 we will apply $\mtbar(\mu)$ with
different choices of $\mu$ for the $X_t$--independent and
$X_t$--dependent contributions to $\mhl^2$,
which will prove useful for absorbing numerically large two-loop contributions
into an effective one-loop result.  In the spirit of EFT, we will argue
that for $\mu=\ms$, one should use the results of \eq{eq:mtsusy} while
for $\mu<\ms$, one should use \eq{eq:mtsmreg}.

A remark on the regularization scheme is
in order here.  In effective field theory, the running top-quark mass
at scales below $\ms$ is the SM running coupling
[\eq{eq:mtsmreg}], which
is calculated in dimensional regularization.
This is matched onto the running top-quark mass as computed in the
full supersymmetric theory.
One could argue that the appropriate regularization
scheme for the latter should be dimensional reduction
(DRED)~\cite{dred}, which is usually applied in loop calculations in
supersymmetry.\footnote{In order
to obtain the corresponding DRED result, one simply has to replace the
term $4 \alpha_s/3 \pi$ in the denominator of
\eq{eq:mtmsbar} by $5 \alpha_s/3 \pi$.}
The result of such a change would be to modify slightly the two-loop
non-logarithmic contribution to $\mhl$ that is proportional to
powers of $X_t$.
Of course, the physical Higgs mass is independent of scheme.
One is free to re-express \eqns{eq:mh1ldiagos}{eq:mh2ldiagos}
[which depend on the on-shell parameters $M_t$, $\ms$, $X_t$]
in terms of parameters defined in any other scheme.  In
this paper, we find \msbar--renormalization via DREG to be
the most convenient scheme
for the comparison of the diagrammatic and EFT results for $\mhl$.

%%%%%%%%%%%%%%%%%%%%%%%%%%%%%%%%%%%%%%%%%%%%%%%%%%%%%%%%%%%%%%%%%%%%%%%%%

\section{Comparing the EFT and diagrammatic results}
\label{sec:sec5}

In order to directly compare the two-loop diagrammatic and EFT results,
we must convert from on-shell to \msbar\ parameters.
Inserting \eqns{eq:msms}{eq:xtrelsusylead} into
\eqns{eq:mh1ldiagosmtbar}{eq:mh2ldiagosmtbar}, one finds
\bea
\mhl^{2,{\alpha}} &=&
  \frac{3}{2} \frac{G_F \sqrt{2}}{\pi^2} \mtbar^4 \left\{
  - \ln\left(\frac{\mtbar^2}{\Msbarii} \right)
  + \frac{\Xtbarii}{\Msbarii}
  \left(1 - \frac{1}{12} \frac{\Xtbarii}{\Msbarii} \right)\right\}\,,
\label{eq:MSbarDREG1} \\
\mhl^{2,{\alpha\alpha_s}} &=&
  - 3 \frac{G_F \sqrt{2}}{\pi^2} \frac{\alpha_s}{\pi} \mtbar^4
  \Biggl\{\ln^2 \left(\frac{\mtbar^2}{\Msbarii} \right) +
    \left[\frac{2}{3} - 2 \frac{\Xtbarii}{\Msbarii}
          \left(1 - \frac{1}{12} \frac{\Xtbarii}{\Msbarii} \right) \right]
    \ln\left(\frac{\mtbar^2}{\Msbarii} \right)  \non \\
&& {} + \frac{\Xtbar}{\Msbar} \left(\frac{2}{3} - \frac{7}{9}
   \frac{\Xtbarii}{\Msbarii} + \frac{1}{36}\frac{\Xtbariii}{\Msbariii}
+ \frac{1}{18} \frac{\Xtbariv}{\Msbariv} \right)
   \Biggr\} + {\cal O}\left(\frac{\mtbar}{\Msbar}\right)\,.
\label{eq:MSbarDREG2}
\eea

Comparing \eq{eq:MSbarDREG2} with \eq{eq:mhrg} shows that
the logarithmic contributions of the diagrammatic result expressed in
terms of the \msbar\ parameters $\mtbar$, $\Msbar$, $\Xtbar$ agree with
the logarithmic contributions obtained by the EFT
approach. The differences in the logarithmic terms
observed in the comparison of \eqns{eq:mh1ldiagosmtbar}{eq:mh2ldiagosmtbar}
with \eq{eq:mhrg} have thus been traced to the different renormalization
schemes applied in the respective
calculations. The fact that the logarithmic contributions obtained
within the two approaches agree after a proper rewriting of the
parameters of the stop sector is an important
consistency check of the calculations.  In addition to the
logarithmic contributions, \eq{eq:MSbarDREG2} also
contains non-logarithmic contributions, which are
numerically sizable.

%%%%%%%%%%%%%%%%%%%%%%%%%%%%%%%%%%%%%%%%%%%%%%%%%%%%%%%%%%%%%%
\begin{figure}[htbp]
\begin{center}
\epsfig{figure=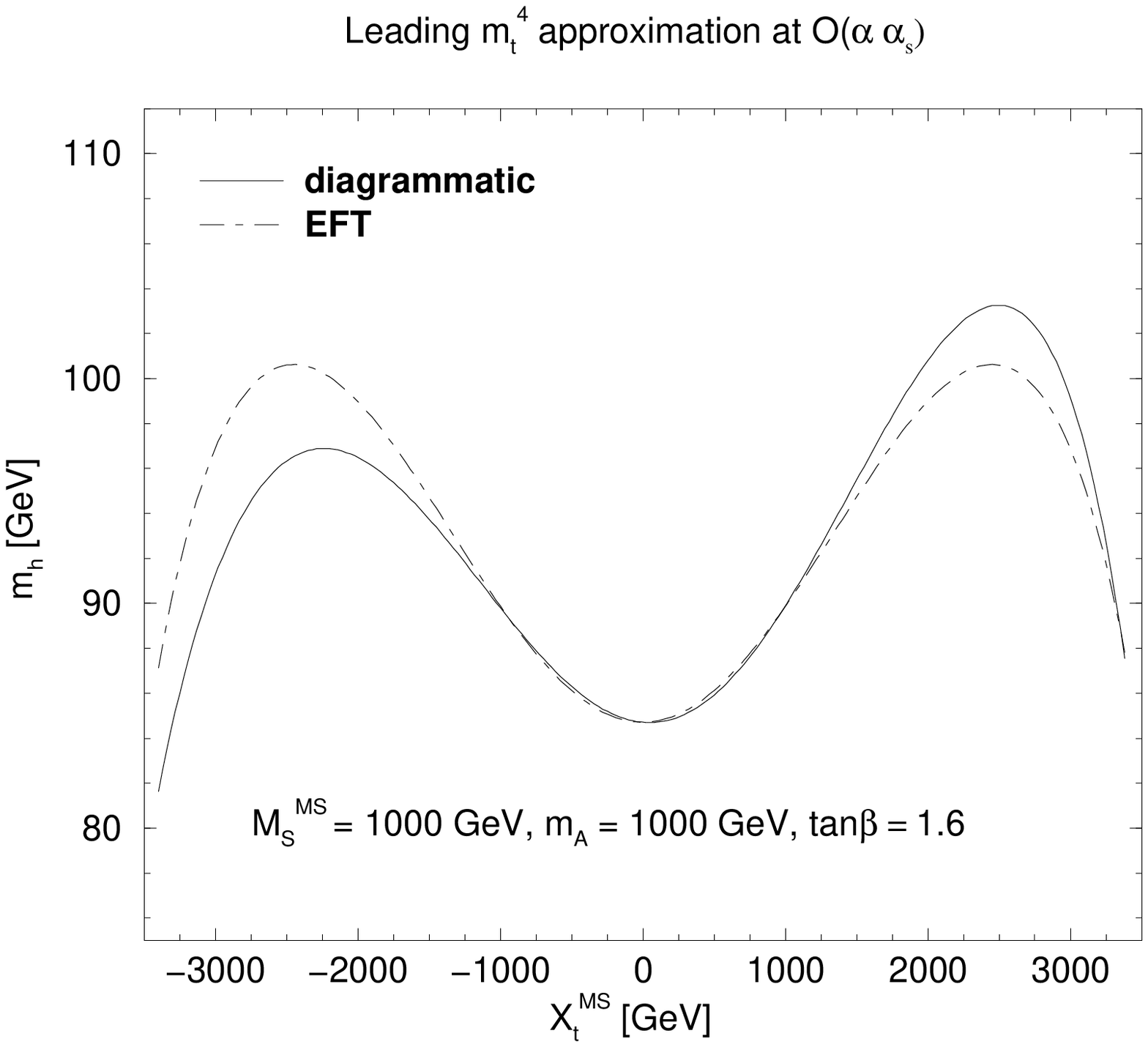,width=12cm,height=9cm}
\end{center}
\vspace{0.2cm}
\begin{center}
\epsfig{figure=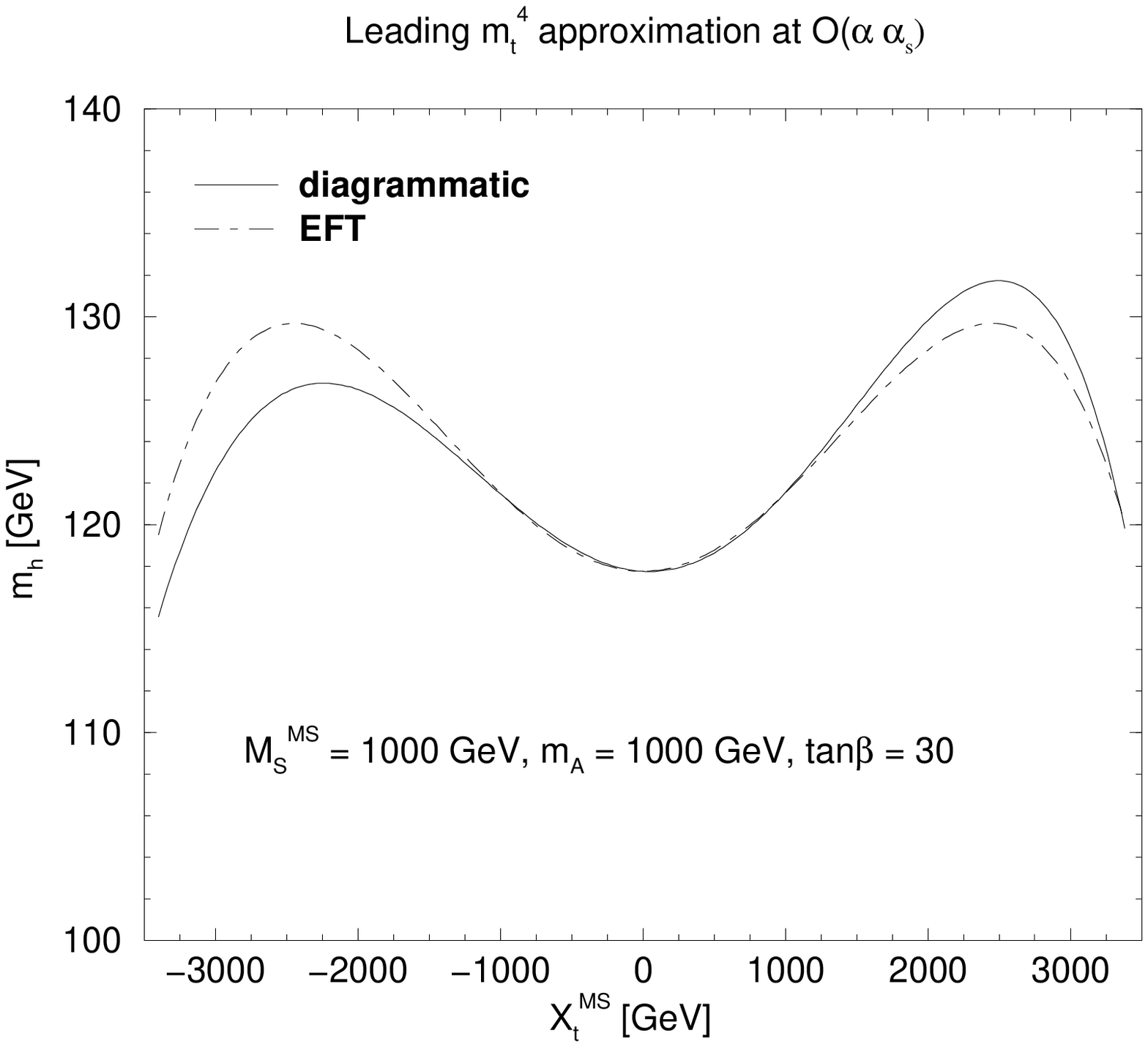,width=12cm,height=9cm}
\end{center}
\fcaption{Comparison of the diagrammatic two-loop
${\cal O}(\mt^2 h_t^2\alpha_s)$ result for $\mhl$, to leading
order in $\mtbar/\Msbar$ [\eqns{eq:MSbarDREG1}{eq:MSbarDREG2}]
with the EFT result of \eq{eq:mhrg}.
Note that the latter omits the
one-loop threshold corrections due to stop mixing in the
evaluation of $\mtbar(\ms)$.  Since this quantity enters in the
definition of $\Xtbar$ [see \eq{eq:xtms}], the meaning of
$X_t^{\rm MS}$ plotted along the $x$-axis is slightly different
for the diagrammatic curve, where $X_t^{\rm MS}=\Xtbar$, and the
EFT curve, where $X_t^{\rm
MS}=\Xtbar\,[1+(\alpha_s/3\pi)(X_t/\ms)]$. See text for further
details. The two graphs above are plotted for
$\Msbar=\mha=(\mgl^2+\mtbar^2)^{1/2}= 1$~TeV for the cases of
$\tanb=1.6$ and $\tanb=30$, respectively. } \label{fig:msbar1}
\end{figure}
%%%%%%%%%%%%%%%%%%%%%%%%%%%%%%%%%%%%%%%%%%%%%%%%%%%%%%%%%%%%%%

In \fig{fig:msbar1}, we compare the diagrammatic result for $\mhl$
in the leading $m_t^4$ approximation to the results obtained in
section 2 by EFT techniques, for two different values of $\tanb$.
However, as noted at the end of section 4, in the derivation of
the EFT result of \eq{eq:mhrg} the supersymmetric threshold
corrections to $\mtbar(\ms)$ were neglected. Thus, $\Xtbar$, which
appears in \eqns{eq:MSbarDREG1}{eq:MSbarDREG2}, is not precisely
the same as the $\Xtbar$ parameter appearing in the EFT result of
\eq{eq:mhrg} due to the difference in the definition of
$\mtbar(\ms)$ [\eq{eq:mtsusy}] and $\mtsmreg$ [\eq{eq:mtsmreg}].
Taking this difference into account in \eq{eq:xtms}, it follows
that the $\Xtbar$ parameter that appears in \eq{eq:mhrg} is given
by $\Xtbarp\equiv \Xtbar\,[1+(\alpha_s/3\pi)(X_t/\ms)]$.
It can be easily checked that
the change from $\Xtbar$ to $\Xtbarp$ does not affect the
comparison of two-loop logarithmic terms between
\eqns{eq:mhrg}{eq:MSbarDREG2}. Moreover, the difference between
$\Xtbar$ and $\Xtbarp$ is numerically small. In \fig{fig:msbar1},
the diagrammatic result for $\mhl$ is plotted versus $\Xtbar$,
while the EFT result is plotted versus $\Xtbarp$.

While the diagrammatic result expressed in terms of $\mtbar$, $\Msbar$,
$\Xtbar$ agrees well with the EFT result in the region of no mixing in
the stop sector, sizable deviations occur for large mixing.
In particular, the non-logarithmic contributions give rise
to an asymmetry under the change of sign of the parameter $\Xtbar$,
while the EFT result is symmetric under $\Xtbar\to -\Xtbar$. In the
approximation considered here, the maximal value for $\mhl$ in
the diagrammatic result lies about 3~GeV higher than the maximal value
of the EFT result for $\tan\beta = 1.6$. The differences are slightly
smaller for $\tan\beta = 30$.  In addition, as previously noted, the
maximal-mixing point $(\Xtbar)_{\rm max}$
[where the radiatively corrected value of $\mhl$ is maximal]
is equal to its one-loop value, $(\Xtbar)_{\rm max}\simeq
\pm\sqrt{6}\ms$, in the EFT result of \eq{eq:mhrg},
while it is shifted in the two-loop diagrammatic result.
However, \fig{fig:msbar1} illustrates that the shift in $(\Xtbar)_{\rm
max}$ from its one-loop value, while significant in the two-loop
on-shell diagrammatic result, is largely diminished when the latter is
re-expressed in terms of \msbar\ parameters.

The differences between the diagrammatic and EFT results shown
in \fig{fig:msbar1} can be attributed to
non-negligible non-logarithmic terms proportional to powers of $X_t$.
Clearly, the EFT technique can be improved to incorporate these terms.
As previously discussed, one can account for such terms in the EFT
approach by: (i) including one-loop finite threshold effects in the
definition of $\mtbar(\ms)$, and (ii) including two-loop finite
threshold effects to $\Delta_{\rm th}\lambda$ [\eq{expumbral}]
in the matching condition for
$\lambda(\ms)$.\footnote{An additional
two-loop ${\cal O}(h_t^2\alpha_s)$ correction to $\lambda(\ms)$
can arise because the Higgs self-coupling in the \msbar\ scheme
does not precisely satisfy the supersymmetric relation
[$\lambda(\ms)=\quarter(g^2+g'^2)\cos^2 2\beta$] in the supersymmetric
limit. This correction corresponds
to a matching of the \msbar\ and \drbar\ couplings at the scale
$\ms$~\cite{radcor}.  We will count this as part of
the two-loop finite threshold effects.}
Although both effects are nominally of the same order, it is
interesting to investigate what fraction of the
terms proportional to powers of $\Xtbar$ in
\eq{eq:MSbarDREG2} can be obtained by including the
threshold effects in the definition of $\mtbar(\ms)$.

In \Ref{mhiggsrg4}, it was shown that the leading two-loop contributions
to $\mhl^2$ given by the EFT result of \eq{eq:mhrg} could be absorbed
into an effective one-loop expression.
This was accomplished by considering separately the
$X_t$--independent leading double
logarithmic term (the ``no-mixing'' contribution) and the
leading single logarithmic term that is proportional to powers of
$\Xtbar$
(the ``mixing'' contribution) at ${\cal O}(m_t^2 h_t^2\alpha_s)$.
Both terms can be reproduced by an effective one-loop expression, where
$\mtbar$ in \eq{eq:MSbarDREG1}, which appears in the
no-mixing and mixing contributions, is replaced by the
{\it running} top-quark mass evaluated at the scales $\mu_t$ and
$\mu_{\tilde t}$, respectively:
\be
\mbox{no mixing: } \mu_t \equiv (\mtbar \Msbar)^{1/2}\,, \qquad
\mbox{mixing: } \mu_{\tilde t} \equiv \Msbar\,.
\label{eq:miximpr}
\ee
That is, at ${\cal O}(m_t^2
h_t^2\alpha_s)$, the leading double logarithmic term
is precisely reproduced by the single-logarithmic term at
${\cal O}(m_t^2 h_t^2)$, by replacing $\mtbar$ with $\mtbar(\mu_t)$,
while the leading single logarithmic term at two-loops proportional to
powers of $\Xtbar$ is
precisely reproduced by the corresponding non-logarithmic terms
proportional to
$\Xtbar$ at ${\cal O}(m_t^2 h_t^2)$, by replacing $\mtbar$ with
$\mtbar(\Msbar)$.

Applying the same procedure to \eq{eq:MSbarDREG2} and rewriting it in
terms of the running top-quark
mass at the corresponding scales as specified in \eq{eq:miximpr},
we obtain
\bea
\mhl^{2,{\alpha}} &=&
  \frac{3}{2} \frac{G_F \sqrt{2}}{\pi^2} \left\{
- \mtbar^4(\mu_t)\ln\left(\frac{\mtbar^2(\mu_t)}{\Msbarii}\right)
+ \mtbar^4(\Msbar)
  \frac{\Xtbarii}{\Msbarii}
  \left(1 -  \frac{\Xtbarii}{12\Msbarii} \right)
  \right\} , % + {\cal O}\left(\frac{\mtbar}{\ms}\right)\,,
  \label{eq:mh1lmixed} \\
\mhl^{2,{\alpha\alpha_s}} &=&
  - 3 \frac{G_F \sqrt{2}}{\pi^2} \frac{\alpha_s}{\pi} \mt^4
  \left\{\frac{1}{6} \ln\left(\frac{\mt^2}{\ms^2} \right) +
  \frac{\Xt}{\ms} \left(\frac{2}{3} - \frac{1}{9} \frac{\Xt^2}{\ms^2} +
  \frac{1}{36} \frac{\Xt^3}{\ms^3} \right)
   \right\} \,.
  %+ {\cal O}\left(\frac{\mt}{\ms}\right).
\label{eq:impr}
\eea
Indeed, the $X_t$--independent leading double logarithmic term
and the leading single logarithmic term that is
proportional to powers of $\Xtbar$ have disappeared from the
two-loop expression [\eq{eq:impr}], having been absorbed into an
effective one-loop result [\eq{eq:mh1lmixed}] (denoted henceforth
as the ``mixed-scale'' one-loop EFT result).  Of the terms that
remain [\eq{eq:impr}], there is a subleading one-loop logarithm at
two-loops which is a remnant of the no-mixing contribution.  But,
note that the magnitude of the coefficient ($1/6$) has been
reduced from the corresponding coefficients that appear in
\eqns{eq:mh2ldiagos}{eq:MSbarDREG2} [$-2$ and $2/3$,
respectively]. In addition, the remaining leftover two-loop
non-logarithmic terms are also numerically insignificant. We
conclude that the ``mixed-scale'' one-loop EFT result provides a
very good approximation to $\mhl^2$, in which the most significant
two-loop terms have been absorbed into an effective one-loop
expression.

%%%%%%%%%%%%%%%%%%%%%%%%%%%%%%%%%%%%%%%%%%%%%%%%%%%%%%%%%%%%%%
\begin{figure}[htbp]
\begin{center}
\epsfig{figure=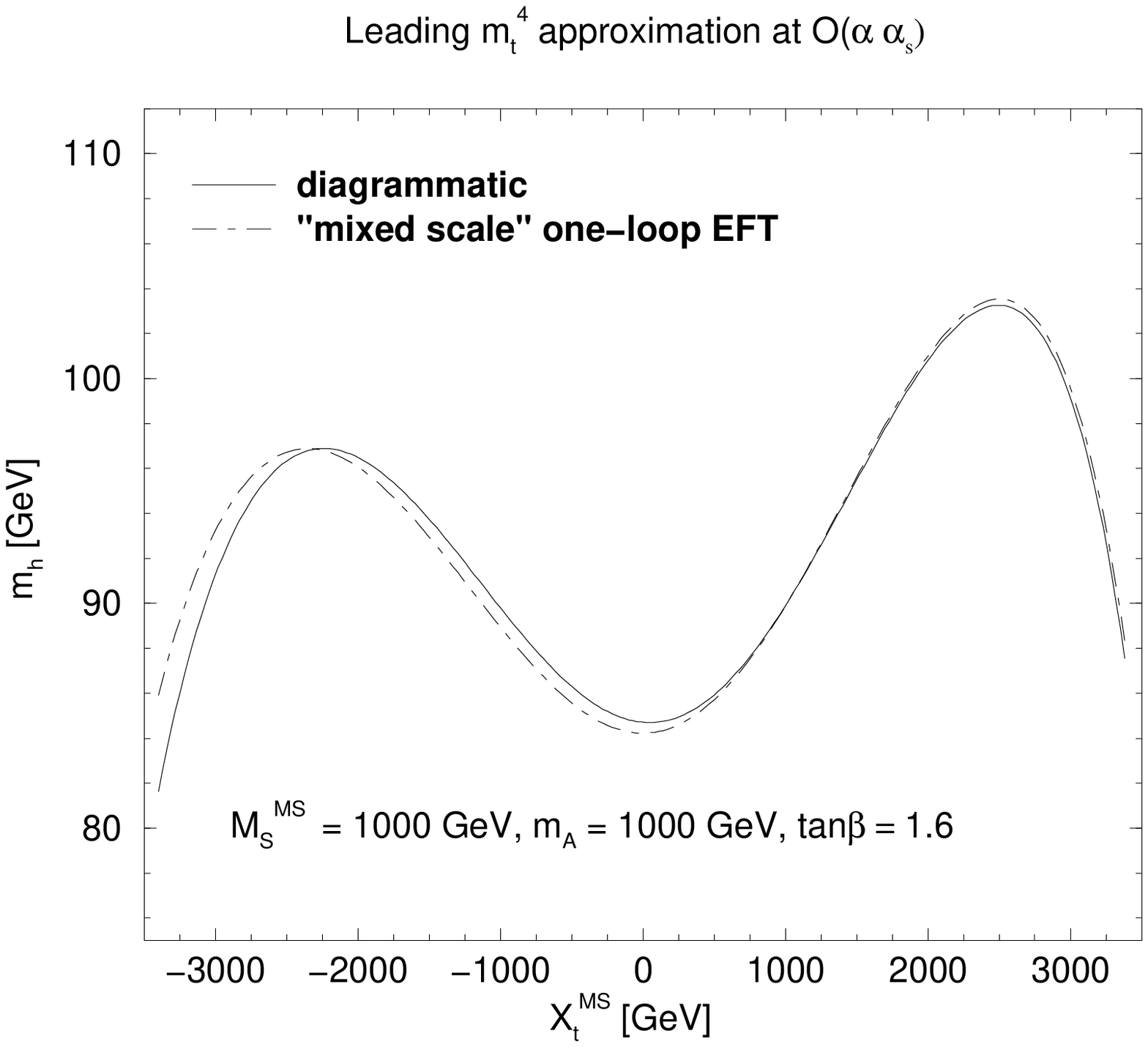,width=12cm,height=9cm}
\end{center}
\vspace{0.2cm}
\begin{center}
\epsfig{figure=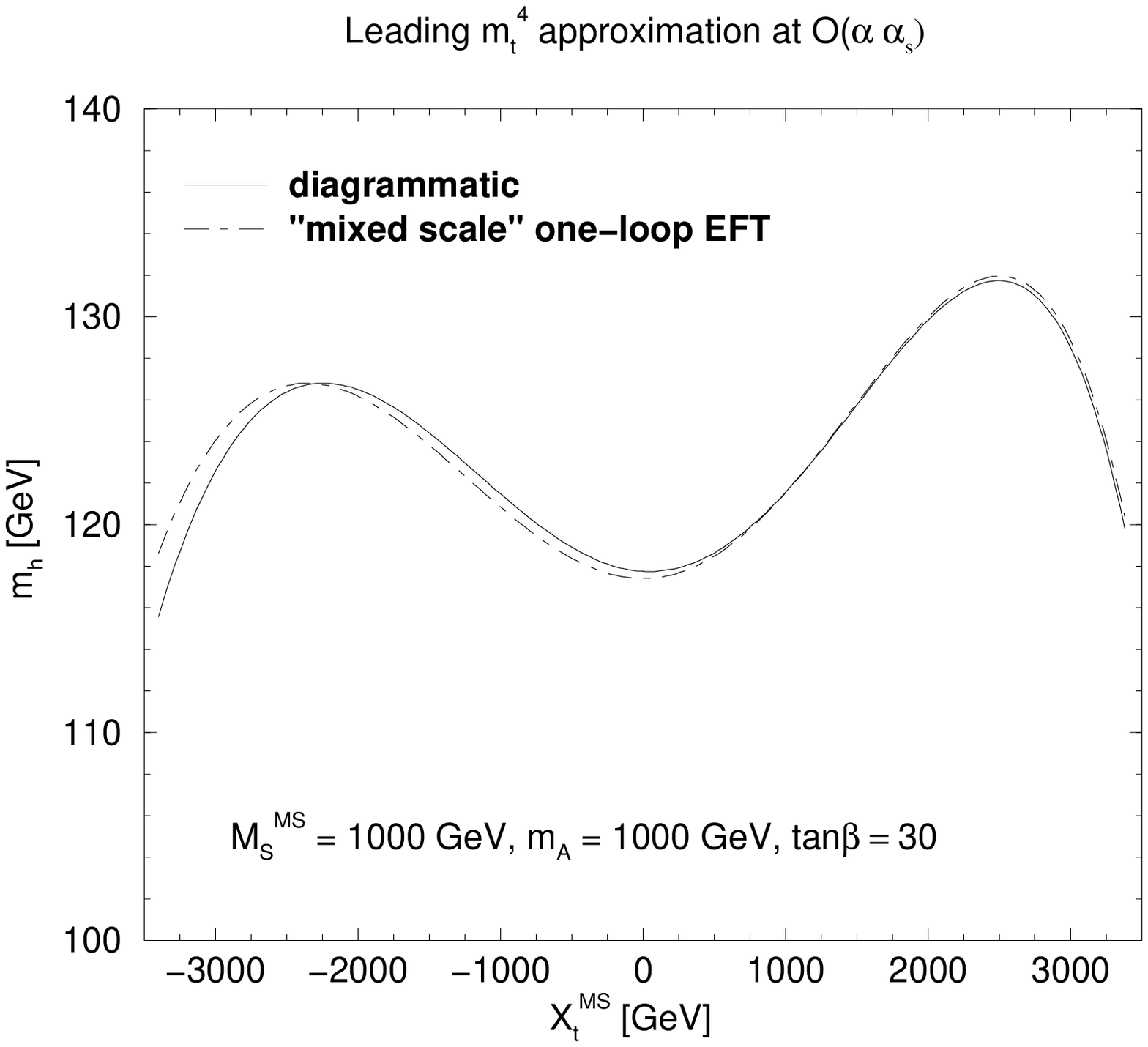,width=12cm,height=9cm}
\end{center}
\fcaption{Comparison of the diagrammatic two-loop
${\cal O}(\mt^2 h_t^2\alpha_s)$ result for $\mhl$, to leading
order in $\mtbar/\Msbar$ [\eqns{eq:MSbarDREG1}{eq:MSbarDREG2}]
with the ``mixed-scale'' one-loop EFT result
[\eq{eq:mh1lmixed}].  Note that the latter now
includes the threshold corrections due to stop mixing in the
evaluation of $\mtbar(\ms)$ in contrast to the EFT results
depicted in \fig{fig:msbar1}.  ``Mixed-scale'' indicates that in
the no-mixing and mixing contributions to the one-loop Higgs mass,
the running top quark mass is evaluated at different scales
according to \eq{eq:miximpr}. See text for further details.  The
two graphs above are plotted for
$\Msbar=\mha=(\mgl^2+\mtbar^2)^{1/2}= 1$~TeV for the cases of
$\tanb=1.6$ and $\tanb=30$, respectively.}
\label{fig:impr1}
\end{figure}
%%%%%%%%%%%%%%%%%%%%%%%%%%%%%%%%%%%%%%%%%%%%%%%%%%%%%%%%%%%%%%

To illustrate this result, we compare in
\fig{fig:impr1} the diagrammatic two-loop result
expressed in terms of \msbar\ parameters
[\eqns{eq:MSbarDREG1}{eq:MSbarDREG2}] with the
``mixed-scale'' one-loop EFT result [\eq{eq:mh1lmixed}]
as a function of $\Xtbar$.  In order to make a fair comparison of
two-loop expressions, we first evaluate \eq{eq:mh1lmixed}
as a perturbation expansion which is truncated
beyond the ${\cal O}(\alpha_s^2)$ term.\footnote{For example, using
\eq{eq:mtsmreg}, we would write $\mtbar^4(\mu_t)=\mtbar^4
\left[1+4(\alpha_s/\pi)\ln(m_t^2/\mu_t^2)\right]$ and insert this into
\eq{eq:mh1lmixed}.}  It is this result that is plotted as a dashed
line in \fig{fig:impr1}.  Note that by construction, the sum of the
two-loop truncated version of \eq{eq:mh1lmixed} and the
leftover two-loop term given by \eq{eq:impr} is equal to the
sum of \eqns{eq:MSbarDREG1}{eq:MSbarDREG2}.  That is, the difference
between the solid and dashed lines of \fig{fig:impr1} is precisely
equal to the leftover two-loop term given by \eq{eq:impr}, which
is seen to be numerically small.
Hence, within the simplifying framework under consideration
({\it i.e.}, only leading $t$--$\Stop$ sector-contributions are taken
into account assuming a simplified stop squared-mass
matrix [\eq{stopmassmatrix}], with
$\Msbar$, $\mha\gg\mtbar$ and $\mgl = \msq$), we see that the
``mixed-scale'' one-loop result for $\mhl$ provides a very good
approximation to a more complete two-loop result for all values of
$\Xtbar$.\footnote{Strictly speaking, the analytic approximations of
this paper break down when $\mtbar\Xtbar\sim\Msbarii$.  Thus, one does
not expect an accurate result for the corresponding formulae when
$\Xtbar$ is too large \cite{mhiggsrg2,mhiggsrg4,mhdiagcomp}.  In
practice, one should not trust the accuracy of the analytic formulae
once $\Xtbar>(\Xtbar)_{\rm max}$.}
In the EFT picture this means that, once
re-expressed in terms of the appropriate $\MS$ running parameters, the
dominant contributions to the lightest $\cp$-even Higgs mass
arising from the two-loop threshold corrections induced by the
decoupling of the stops, have their origin in the one-loop
threshold corrections to the Higgs--top quark Yukawa coupling.

In the present analysis we have focused on the leading
contributions of the $t$--$\Stop$ sector of the MSSM.  However,
these contributions alone are not sufficient to provide an
accurate determination of the Higgs mass (and can be off by 5~GeV
or more in certain regions of the MSSM parameter space). In
any realistic phenomenological analysis of the properties of the
Higgs sector, one must include sub-leading contributions of the
$t$--$\Stop$ sector as well as contributions from other
particle/superpartner sectors. Such contributions have been
obtained within the EFT approach in \Refs{mhiggsrg3}{mhiggsrg4},
which incorporates one-loop leading logarithmic terms from all
partner/superpartner sectors, plus single and double logarithmic
two-loop contributions from the $t$--$\Stop$ and
$b$--$\widetilde{b}$ sectors. The full one-loop diagrammatic
result is known~\cite{fulloneloop1,mhiggsOL2,fulloneloop2}, and
this along with the diagrammatic two-loop result from the
$t$--$\Stop$ sector at ${\cal O}(\alpha\alpha_s)$ are included in
the Fortran code \fh~\cite{feynhiggs}. The impact of the
additional contributions to the radiatively-corrected Higgs mass
on phenomenological studies have been investigated in
refs.~\cite{mhiggsdiag3,tbexcl}.

In the large $\tan\beta$ regime, the $b$--$\widetilde{b}$ sector
is especially important. Here the corrections induced by the
bottom Yukawa coupling become relevant, and one should
correspondingly include the bottom mass corrections originating
from the decoupling of the supersymmetric particles. These
corrections are enhanced by a large $\tan\beta$ factor and hence
can have a sizable impact on the phenomenology of the Higgs
sector. Some of the most relevant consequences of these
corrections have been recently discussed in
refs.~\cite{steve1,babu,pilaftsis}.

%%%%%%%%%%%%%%%%%%%%%%%%%%%%%%%%%%%%%%%%%%%%%%%%%%%%%%%%%%%%%%%%%%%%%%%%%

\section{Discussion and Conclusions}

In this work, we have compared the results for the lightest $\cp$-even
Higgs-boson mass obtained from the two-loop  ${\cal O}(\alpha\alpha_s)$
diagrammatic calculation in the on-shell scheme with the
results of an effective field theory approach.  In the latter, the
two-loop ${\cal O}(\alpha\alpha_s)$ terms are generated via renormalization
group running of the effective low-energy parameters from the
supersymmetry-breaking scale, $\ms$ to the scale $\mt$.
We have focused on the leading ${\cal O}(m_t^2 h_t^2\alpha_s)$
two-loop contributions to $\mhl^2$
in the limit of large $\mha$ and $M_S$.
In this case, the effective field theory below $M_S$ is the
one-Higgs-doublet Standard Model, which greatly simplifies the
calculation.
In addition, the gluino mass was set to $\mgl = \msq\equiv
(\ms^2-\mt^2)^{1/2}$.
The resulting transparent analytic expressions for the
radiatively-corrected Higgs mass were
well suited for investigating the basic relations
between the various approaches.  In order to compare the on-shell
diagrammatic and effective field theory approaches,
one must note two important facts.  First, the two calculations are
performed in different renormalization schemes. Hence, the resulting
expressions actually depend on soft-supersymmetry-breaking parameters
whose definitions differ at the one-loop level.  Second,
the diagrammatic calculation includes genuine non-logarithmic \twol\
corrections to the lightest $\cp$-even Higgs-boson mass.  In the
effective field theory approach, these would correspond to
two-loop threshold corrections resulting from the
decoupling of the two heavy top squarks in the low energy effective
theory.

Previous comparisons of the corresponding two-loop
results revealed an apparent discrepancy between terms that depend
logarithmically on $\ms$, and
different dependences of the non-logarithmic terms on
$X_t$.    However, after
re-expressing the \onel\ and the \twol\ terms of the on-shell
diagrammatic result in terms of \msbar\ parameters,
{\it i.e.}, applying the same renormalization scheme for both
approaches, we have shown that the discrepancy in the logarithmic
dependence on $\ms$ of
both expressions disappears. This constitutes an important consistency
check of the calculations. There remain, however, genuine
non-logarithmic \twol\ contributions in the diagrammatic result. They
give rise to an asymmetry under $\Xt\to -\Xt$, while the
effective field theory computations that neglected the
two-loop threshold corrections due to stop mixing only yield results
that are symmetric under the change of sign of $\Xt$.  Moreover,
the non-logarithmic \twol\ contributions of the on-shell diagrammatic
computation, in the approximations considered in this paper,
can {\it increase}
the predicted value of $\mhl$ by as much as 3~GeV.\footnote{If only the
on-shell
top-quark mass is re-expressed in terms of the corresponding \msbar\
parameter, the resulting increase in $\mhl$ can be as much as 5 GeV.}
Finally, they induce a shift of the value of $\Xt$ where
$\mhl$ is maximal relative to the corresponding one-loop value
$(X_t)_{\rm max}\simeq \pm\sqrt{6}\ms$.  It is interesting to note that
this shift is
more (less) pronounced when $\mhl$ is expressed in terms of $\Xt$ in the
on-shell (\msbar) scheme.  The effect of the leading non-logarithmic
\twol\ contributions can be taken into account in the effective field
theory method by incorporating the ${\cal O}(h_t^2\alpha_s)$
$\Xt$--dependent corrections
into the boundary conditions of the effective quartic Higgs
self-coupling at
the scale $\ms$ and performing a proper one-loop ${\cal
O}(\alpha_s)$ matching of the running Higgs--top Yukawa coupling at
$\ms$.

In \Ref{mhiggsrg4}, it was shown that the leading two-loop contributions
to $\mhl^2$ could be absorbed into an effective one-loop expression by
the following procedure.  The running top-quark mass that appears in the
$X_t$--independent one-loop expression for $\mhl^2$ is evaluated
at the scale $\mu_t=(\ms\mt)^{1/2}$.  In the corresponding
$X_t$--dependent terms, the running top quark mass is evaluated
at the scale $\mu_{\tilde t}=\ms$.
The result, which we call the {\it
mixed-scale} one-loop EFT expression neatly incorporates the leading
two-loop effects.  In this paper, we have extended this result by
explicitly including stop mixing effects in evaluating the running
parameters $\mtbar(\ms)$ and $h_t(\ms)$.  By doing so, we are able to
incorporate some portion of the genuine leading non-logarithmic
\twol\ contributions.  The remaining terms at this order would then be
identified with two-loop threshold corrections to the effective Higgs
quartic coupling at the scale $\ms$.  Remarkably, the latter turn out to
be numerically small.  This means that the mixed-scale one-loop EFT
expression for $\mhl$ provides a rather accurate estimate of the
radiatively-corrected mass of the lightest $\cp$-even Higgs boson of the
MSSM.

The above results have been obtained in a rather simple setting.  A
special choice for the stop squared-mass matrix was made
[\eq{stopmassmatrix}] to simplify
our analysis.  The gluino mass was fixed to a value of order $\ms$.
The leading ${\cal O}(m_t^2 h_t^4)$ corrections were neglected.
Subleading terms of ${\cal O}(\mz^2 h_t^2\alpha_s)$ and
${\cal O}(\mz^2 h_t^4)$ terms were also neglected.  For example,
consider the effect of varying the gluino mass.  The two-loop
diagrammatic results of
refs.~\cite{mhiggsdiag1,mhiggsdiag2,mhiggsdiag3,mhdiagcomp} showed that
the value of $\mhl$ changed by as much as $\pm 2$~GeV as a function of
$\mgl$, for $\mt\lsim\mgl\lsim\ms$.  The gluino mass dependence can be
treated in the EFT approach as follows.  Let us assume that $\ms$
characterizes the scale of the squark masses, and $\mgl<\ms$.  Then,
at scales below $\ms$ one integrates out the squarks but keeps the
gluino as part of the low-energy effective theory.  However, since the
gluino {\it always} appears with squarks in diagrams contributing to
$\mhl^2$ at two-loops, once the squarks are integrated out, they no
longer affect the running of any of the relevant low-energy parameters
below $\ms$.  However, the gluino mass does affect the value of
$h_t(\ms)$ and $\mtbar(\ms)$ [the relevant formulae are given in
Appendix~A].  Thus, in the EFT approach, gluino mass dependence enters
via the threshold corrections to the Higgs--top quark Yukawa couplings.

The case of a more general stop squared-mass matrix can be treated using
the same methods outlined in Appendix~A.\footnote{The transformation
from on-shell to \msbar\ input parameters for the case of the most
general stop squared-mass matrix will be included in the new version of
the program \fh.}   Here the computations are more
complicated since there is now one-loop mixing between $\widetilde t_1$
and $\widetilde t_2$.  In the EFT approach, one must decouple separately
the two stops, and include the most general stop-mixing effects in the
determination of the relation between the on-shell and \msbar\
parameters. The leading ${\cal O}(m_t^2 h_t^4)$ corrections in the
EFT approach can be incorporated as in
section~2~\cite{mhiggsrg3,mhiggsrg4}
by extending the computations of Appendix~A to include
the one-loop ${\cal O}(h_t^2)$ corrections to the running top-quark
mass and stop sector parameters.  However, at present, one cannot check
these results against an ${\cal O}(m_t^2 h_t^4)$
two-loop diagrammatic computation, since the latter does not yet appear
in the literature in full generality.

Going beyond the approximations made in this paper, the next step is
to incorporate the above improvements, as well as the next subleading
contributions of
${\cal O}(\mz^2 h_t^2\alpha_s)$ and ${\cal O}(\mz^2 h_t^4)$
into the computation of $\mhl^2$.
One might then hope to show that a complete mixed-scale one-loop EFT
result, suitably generalized, provides a very good approximation to the
radiatively-corrected
$\cp$-even Higgs mass of the MSSM.  Such an analysis could be used to
organize the most significant
non-leading one-loop and two-loop contributions to $\mhl^2$
and provide some insight regarding the magnitude of
the unknown higher-order corrections,
thus reducing the theoretical
uncertainty in the prediction for $\mhl$. This would have a significant
impact on the physics of the lightest \cp-even Higgs boson at LEP2, the
upgraded Tevatron and the LHC.

\newpage
%\vskip1cm
\noindent{\large\bf Note Added} \vskip 0.4cm

\noindent
After this work was completed, we received a paper~\cite{espizh}
which discusses many of the same issues that we address
in this work.  In \Ref{espizh}, the two-loop effective
potential of the MSSM is employed,
including renormalization group resummation of logarithmic terms, and
the leading non-logarithmic two-loop terms of ${\cal
O}(\alpha\alpha_s)$.  The relation between on-shell and \msbar\
parameters are also taken into account.  The end results are
qualitatively similar to the ones obtained here.

\vskip1cm \noindent{\large\bf Acknowledgments} \vskip 0.4cm

\noindent
M.C., H.E.H. and C.W. would like to thank the Aspen
Center for Physics, where part of this work has been carried out.
The work of M.C., H.E.H. and C.W. is supported in part by the
U.S. Department of Energy. W.H. gratefully acknowledges support by
the Volkswagenstiftung.  H.E.H. and S.H. thankfully acknowledge
hospitality of the CERN Theory Group, where a major part of this
work was completed. H.E.H. is pleased to acknowledge useful
discussions with Damien Pierce.

%\newpage
\vskip 0.4in

\appendixA{Appendix~A: \boldmath{Relations between on-shell
and $\overline{\rm MS}$ definitions of $\mt$, $\ms$ and $X_t$}}

In this appendix, we derive the relations between the on-shell and
\msbar\ definitions of $m_t$, $\ms$ and $X_t$.  We have checked that
these results agree with similar results given in \Ref{fulloneloop2}.
These results are derived
in a model where the stop mass-squared matrix is given by
\eq{stopmassmatrix}.
The corresponding stop squared-masses and mixing angle are given
by \eqns{mstopins}{mixangins}.
Note that in this model, the top-squark mass eigenstates, $\stopa$ and
$\stopb$, are states of definite
parity in their interactions with gluons and gluinos.
The corresponding Feynman rules are shown in \fig{feynman}.
%%%%%%%%%%%%%%%%%%%%%%%%%%%%
\begin{figure}[htbp]
\begin{center}
\begin{picture}(240,86)(0,0)
\Line(50,40)(100,40)
\ArrowLine(100,40)(140,70)
\DashArrowLine(140,10)(100,40)5
\Text(0,40)[]{(a)}
\Text(40,40)[]{$\widetilde g$}
\Text(152,10)[]{$\widetilde t_1$}
\Text(150,70)[]{$t$}
\Text(180,40)[l]{$-ig_s T^a\times
\left\{\begin{array}{ll}
1\,, \mbox{when $X_t>0$} \\[5pt]
\!\!\gammav\,, \mbox{when $X_t<0$} \end{array}\right.$}
\end{picture}
\end{center}
%%%%%%%%%%%%%%%%%%%%%%%%%%%%
\begin{center}
\begin{picture}(240,86)(0,0)
\ArrowLine(50,40)(100,40)
\ArrowLine(100,40)(140,70)
\DashArrowLine(140,10)(100,40){5}
\Text(0,40)[]{(b)}
\Text(40,40)[]{$\widetilde g$}
\Text(152,10)[]{$\widetilde t_2$}
\Text(150,70)[]{$t$}
\Text(180,40)[l]{$-ig_s T^a\times
\left\{\begin{array}{ll}
\gammav\,, \mbox{when $X_t>0$} \\[5pt]
\!\!-1\,, \mbox{when $X_t<0$} \end{array}\right.$}
\end{picture}
\end{center}
%%%%%%%%%%%%%%%%%%%%%%%%%%%%
\begin{center}
\begin{picture}(240,86)(0,0)
\Gluon(50,40)(100,40){5}{4}
\DashArrowLine(140,70)(100,40){5}
\DashArrowLine(100,40)(140,10){5}
\Text(0,40)[]{(c)}
\Text(40,40)[]{$g$}
\Text(120,15)[]{$p'$}
\Text(120,65)[]{$p$}
\Text(152,70)[]{$\widetilde t_i$}
\Text(152,10)[]{$\widetilde t_j$}
\Text(180,40)[l]{$-ig_s T^a (p+p')_\mu \delta_{ij}$}
\end{picture}
\end{center}
%%%%%%%%%%%%%%%%%%%%%
\begin{center}
\begin{picture}(240,86)(0,0)
\DashArrowLine(60,10)(100,40){5}
\DashArrowLine(60,70)(100,40){5}
\DashArrowLine(100,40)(140,70){5}
\DashArrowLine(100,40)(140,10){5}
\Vertex(100,40){2}
\Text(0,40)[]{(d)}
\Text(52,10)[]{$\widetilde t_j$}
\Text(52,70)[]{$\widetilde t_j$}
\Text(152,70)[]{$\widetilde t_j$}
\Text(152,10)[]{$\widetilde t_j$}
\Text(80,14)[]{$k$}
\Text(80,66)[]{$n$}
\Text(120,14)[]{$\ell$}
\Text(120,66)[]{$m$}
\Text(180,40)[l]{$-ig_s^2(T^a_{\ell k}T^a_{mn}+T^a_{\ell n}T^a_{mk})$}
\end{picture}
\end{center}
%%%%%%%%%%%%%%%%%%%%%
\begin{center}
\begin{picture}(240,86)(0,0)
\DashArrowLine(60,10)(100,40){5}
\DashArrowLine(60,70)(100,40){5}
\DashArrowLine(100,40)(140,70){5}
\DashArrowLine(100,40)(140,10){5}
\Vertex(100,40){2}
\Text(0,40)[]{(e)}
\Text(52,10)[]{$\widetilde t_2$}
\Text(52,70)[]{$\widetilde t_1$}
\Text(152,70)[]{$\widetilde t_1$}
\Text(152,10)[]{$\widetilde t_2$}
\Text(80,14)[]{$k$}
\Text(80,66)[]{$n$}
\Text(120,14)[]{$\ell$}
\Text(120,66)[]{$m$}
\Text(180,40)[l]{$-ig_s^2T^a_{\ell k}T^a_{mn}$}
\end{picture}
\end{center}
%%%%%%%%%%%%%%%%%%%%%
\fcaption{\label{feynman} Feynman rules for top-squark interactions
in a model where the stop mass-squared matrix is given by
\eq{stopmassmatrix}.}
\end{figure}
%\vspace*{3mm}

%%%%%%%%%%%%%%%%%%%%%%%%%%%%
\begin{figure}[htbp]
\begin{center}
\begin{picture}(400,100)(0,0)
\GlueArc(100,40)(40,0,180){5}{8}
\ArrowLine(60,40)(140,40)
\ArrowLine(10,40)(60,40)
\ArrowLine(140,40)(190,40)
\Text(100,8)[]{(a)}
%\Vertex(60,100){2}
%\Vertex(140,100){2}
%
%
%
\CArc(300,40)(40,0,180)
\DashArrowLine(260,40)(340,40){5}
\ArrowLine(210,40)(260,40)
\ArrowLine(340,40)(390,40)
\Text(235,27)[]{$t$}
\Text(365,27)[]{$t$}
\Text(300,90)[]{$\widetilde g$}
\Text(300,27)[]{$\widetilde t_{1,2}$}
\Text(300,8)[]{(b)}
\end{picture}
\end{center}
%\centerline{Fig.~2. One-loop contributions to the top quark mass.}
\fcaption{\label{qoneloop} One-loop contributions to the top
quark mass.}
\end{figure}
%%%%%%%%%%%%%%%%%%%%%%%%%%%%%%%%%%%%%%%%%
%\vspace*{3mm}

Consider first the one-loop contribution to the top-quark two-point
function at ${\cal O}(\alpha_s)$ due to: (i) the top-quark/gluon loop
[\fig{qoneloop}(a)] and (ii) the stop/gluino loop
[\fig{qoneloop}(b)].  Divergences are regulated
by dimensional regularization in $D\equiv 4-2\epsilon$ dimensions
and removed by minimal subtraction.  Including the
tree-level contribution (which is equal to the negative of the inverse
tree-level propagator), the end result is\footnote{All results in this
section are given in the \msbar\ subtraction scheme.  In the
\drbar scheme, all the formulae of this appendix still apply except in
the case of the top-quark gluon loop.  To obtain the corresponding
\drbar result, simply
remove the additive factors of 1 in the two occurences in \eq{gam2f}.}
\bea  \label{gam2f}
\Gamma^{(2)}(p)&=&i[\pslash-\mtbar(\mu)]-{iC_F\alpha_s\over 4\pi}
\left\{\pslash[1+2\Bone (p^2;m_t^2,0)]
-2m_t[1-2\Bzero (p^2;m_t^2,0)]\right.\nonumber \\
&&
+\pslash[\Bone (p^2;\mgl^2,\ms^2-m_t X_t)+ \Bone (p^2;\mgl^2,\ms^2+m_t
X_t)]\nonumber \\ && \left.
-\mgl[\Bone (p^2;\mgl^2,\ms^2-m_t X_t)- \Bone (p^2;\mgl^2,\ms^2+m_t
X_t)]\right\}\,,
\eea
where $T^a T^a=C_F{\bf 1}$ is the SU(3) quadratic Casimir operator
in the fundamental representation, $C_F=4/3$, and
\be \label{bndef}
\Bn (p^2;m_1^2,m_2^2)\equiv (-1)^{n+1}\int_0^1\,dy\,y^n \,
\ln\left({m_2^2y+m_1^2(1-y)-p^2 y(1-y)\over\mu^2}\right)\,.
\ee
The $\Bn$ ($n=0$,1) are related to the standard two-point loop
functions that arise in one-loop computations~\cite{hollik}:
\bea
B_0(p^2;m_1^2,m_2^2)&\equiv &\Delta +\Bzero (p^2;m_1^2,m_2^2)
\,,\nonumber \\
B_1(p^2;m_1^2,m_2^2)&\equiv &-\half\Delta + \Bone (p^2;m_1^2,m_2^2)\,,
\eea
where all occurrences of $\Delta\equiv (4\pi)^\epsilon \Gamma(\epsilon)$
are removed in the minimal subtraction procedure.  Note that
$B_0(p^2;m_1^2,m_2^2)$ is invariant under the interchange of $m_1^2$ and
$m_2^2$, whereas
\be \label{boneswitch}
B_1(p^2;m_2^2,m_1^2)=-B_1(p^2;m_1^2,m_2^2)-B_0(p^2;m_1^2,m_2^2)\,.
\ee
In \eq{gam2f}, $\mu$ is the arbitrary mass parameter of the
\msbar--scheme.
%although no physical quantity can depend on it.
The on-shell top-quark mass, $M_t$, is defined by
$\Gamma^{(2)}
(\pslash=M_t)=0$. It follows that
\bea
M_t &=& \mtbar(\mu)+{C_F\alpha_s m_t\over 4\pi}\Biggl\{
4\Bzero(m_t^2;m_t^2,0)+2\Bone(m_t^2; m_t^2,0) -1 \nonumber
\\
&+& \Bone(m_t^2;\mgl^2,\ms^2-\mt X_t)+
\Bone(m_t^2;\mgl^2,\ms^2+\mt X_t) \nonumber  \\
&-& {\mgl\over m_t}\left[\Bzero(m_t^2;\mgl^2,\ms^2-\mt X_t)
-\Bzero(m_t^2;\mgl^2,\ms^2+\mt X_t)\right]\Biggr\}\,.
\label{mtos}
\eea
In the
${\cal O}(\alpha_s)$ terms above, we simply use the
generic notation $\mt$
for the top-quark mass, since to one-loop accuracy one need not
distinguish between $\alpha_s M_t$ and $\alpha_s\mt(\mu)$.
As previously noted, the relation between the top-quark
mass in the on-shell and \drbar schemes is obtained by dropping the
$-1$ (which does not multiply a loop-function) in \eq{mtos}.

Two of the loop functions are easily evaluated: $\Bzero (m_t^2;
m_t^2,0)=2-\ln(\mt^2/\mu^2)$ and $\Bone (m_t^2;
m_t^2,0)=-\half[3-\ln(\mt^2/\mu^2)]$. We note the following
curious fact.  In the supersymmetric limit ($\mstopa=\mstopb=m_t$
and $\mgl=0$) at one-loop, the relation between the on-shell and
\drbar running top-quark mass evaluated at $\mu=\mt$ is precisely the
same as the corresponding relation between the on-shell and \msbar\
running top-quark mass in non-supersymmetric QCD.

%%%%%%%%%%%%%%%%%%%%%%%%%%%%
\begin{figure}[htbp]
\begin{center}
\begin{picture}(420,120)(0,0)
\CArc(90,70)(40,0,180)
\ArrowArc(90,70)(40,180,360)
\DashArrowLine(0,70)(50,70){5}
\DashArrowLine(130,70)(180,70){5}
\Text(25,56)[]{$\widetilde t_i$}
\Text(155,56)[]{$\widetilde t_j$}
\Text(90,120)[]{$\widetilde g$}
\Text(90,20)[]{$t$}
\Text(90,0)[]{(a)}
%\Vertex(60,100){2}
%\Vertex(140,100){2}
%
%
%
\GlueArc(310,70)(40,0,180){5}{8}
\DashArrowLine(270,70)(350,70){5}
\DashArrowLine(220,70)(270,70){5}
\DashArrowLine(350,70)(400,70){5}
\Text(245,56)[]{$\widetilde t_j$}
\Text(375,56)[]{$\widetilde t_j$}
\Text(310,20)[]{(b)}
\end{picture}
%\vspace*{30mm}
\begin{picture}(400,180)(0,0)
\GlueArc(90,100)(36,-90,270){5}{13}
\DashArrowLine(0,60)(90,60){5}
\DashArrowLine(90,60)(180,60){5}
\Vertex(90,60){2}
\Text(30,46)[]{$\widetilde t_j$}
\Text(145,46)[]{$\widetilde t_j$}
\Text(90,30)[]{(c)}
%\Vertex(60,100){2}
%\Vertex(140,100){2}
%
%
%
\DashArrowArc(310,100)(39,-90,270){5}
\DashArrowLine(220,60)(310,60){5}
\DashArrowLine(310,60)(400,60){5}
\Vertex(310,60){2}
\Text(250,46)[]{$\widetilde t_j$}
\Text(365,46)[]{$\widetilde t_j$}
\Text(310,155)[]{$\widetilde t_i$}
\Text(310,30)[]{(d)}

\end{picture}
\end{center}
%\centerline{Fig.~3. One-loop contributions to the top squark mass.}
\fcaption{\label{sqoneloop} One-loop contributions to the top
squark mass.}
\end{figure}
%%%%%%%%%%%%%%%%%%%%%%%%%%%%%%%%%%%%%%%%%

Next, we examine the one-loop contributions to the top-squark
two-point function at ${\cal O}(\alpha_s)$.  The contributing graphs are
shown in \fig{sqoneloop}.  We immediately note that graph (c)
of \fig{sqoneloop} vanishes
in dimensional regularization.  Moreover, since $\stopa$ and $\stopb$
are states of definite parity
in a model where the stop mass-squared matrix is given by
\eq{stopmassmatrix}, the one-loop mixing of $\stopa$ and
$\stopb$ vanishes to all orders in $\alpha_s$.  Including the tree-level
contribution, the final result for the top squark two-point
function is\footnote{In deriving \eq{stoptwo}, we note that the
contribution of \fig{sqoneloop}(d) [only the case $i=j$ yields a
non-zero contribution, which is equal to
$(iC_F\alpha_s/4\pi)\Azero(\mstop)$] cancels
a similar term that arises in the evaluation of \fig{sqoneloop}(b).}
\bea \label{stoptwo}
&&\hskip -0.9cm\widetilde\Gamma_{jj}^{(2)}(p^2) =
i(p^2-\mstop^2)-{iC_F\alpha_s\over\pi}
\Bigl[\Azero(\mgl^2)+m_t^2\Bzero(p^2;m_t^2,\mgl^2)\nonumber \\
&& +p^2\Bone(p^2;m_t^2,\mgl^2)-(-1)^j\mgl
m_t\Bzero(p^2;m_t^2,\mgl^2)-p^2\Bone(p^2;\mstop^2,0)\Bigr]\,,
\eea
for $j=1$, 2, where
\be
\Azero(m^2)\equiv m^2\left[1-\ln\left({m^2\over\mu^2}\right)\right]
\ee
is related to the standard one-point loop function
$A_0(m^2)=m^2\Delta+\Azero(m^2)$.

There is no distinction in this calculation between the \msbar\
and \drbar schemes.  The on-shell stop squared-masses are defined by
$\widetilde\Gamma_{jj}^{(2)}(p^2=(\mstop^{\rm OS})^2)=0$. Noting
the form for the tree-level squared-masses [\eq{mstopins}], it
follows that:
\be \label{ostoms}
\ms^{2,{\rm OS}}\mp M_t X_t^{\rm OS}=
\Msbarii(\mu)\mp\mtbar(\mu) \Xtbar(\mu)+{C_F\alpha_s\over\pi}
\left[f(\ms^2\mp\mt X_t)\pm g(\ms^2\mp\mt X_t)\right]\,,
\ee
where $M_t$ is the on-shell top quark mass and
\bea \label{fandg}
f(p^2) &\equiv&
\Azero(\mgl^2)+\mt^2\Bzero(p^2;\mt^2,\mgl^2)+
p^2\Bone(p^2;\mt^2,\mgl^2)\,, \nonumber \\
g(p^2) &\equiv& \mgl\mt\Bzero(p^2;\mt^2,\mgl^2)\,.
\eea
It is then
straightforward to solve for $\ms^{2,{\rm OS}}$ and $M_t
X_t^{\rm OS}$ in terms of the corresponding \msbar\ quantities
evaluated at the scale $\mu=\ms$.  Using the notation of
\eq{eq:MsXtbar}, we obtain
\bea
\ms^{2,{\rm OS}} &=& \Msbarii+{C_F\alpha_s\over
2\pi} \left[f(\ms^2+\mt X_t)+f(\ms^2-\mt X_t)\right. \nonumber \\
&&\qquad\qquad \left. -g(\ms^2+\mt X_t) +g(\ms^2-\mt
X_t)\right]\,, \label{mstos}\\
M_t X_t^{\rm OS} &=& \mtbar(\ms) \Xtbar+{C_F\alpha_s\over
2\pi} \left[f(\ms^2+\mt X_t)-f(\ms^2-\mt X_t)\right.\nonumber \\
&&\qquad\qquad \left. -g(\ms^2+\mt X_t)-g(\ms^2-\mt X_t)\right]\,,
\label{Xtos}
\eea
where all loop functions in \eqns{mstos}{Xtos} are evaluated at
$\mu=\ms$.   Dividing \eq{Xtos} by $M_t$, and using \eq{mtos} to
evaluate $\mtbar(\ms)/M_t$, one obtains a direct relation between
$X_t^{\rm OS}$ and $\Xtbar$.

To obtain the expansions derived in Appendix~B, we consider the case of
$\mgl^2=\msusy^2=\ms^2-\mt^2$.
We introduce the following notation:
\be \label{xtzdef}
x_t\equiv {X_t\over\ms}\,,\qquad z\equiv{M_t\over\ms}\,.
\ee
We are interested in the limit of $z\ll 1$ and $x_t\lsim 1$.  First,
consider the relation between the on-shell and running top quark
mass.  Using \eq{bndef}, we must evaluate
the following integrals:
\be  \label{jint}
J^{(\pm)}_n=\int_0^1\,dy\,y^n \,\ln\left[1\pm x_t zy-z^2(1-y^2)\right]\,,
\ee
for $n=0$, 1.  \Eq{mtos} then yields:
\be  \label{mtmsos}
\mtbar(\ms)=M_t\left\{1+{C_F\alpha_s\over 4\pi}\left[-4+6\ln z
-J^{(+)}_1 -J^{(-)}_1+{1\over
z}(1-z^2)^{1/2}(J^{(+)}_0-J^{(-)}_0)\right]\right\}\,.
\ee
Expanding out the logarithm in the integrand of $J^{(\pm)}_n$
in a double power series in $x_t$ and $z$ and integrating term by term,
one readily obtains the result given in \eq{mtsusymsbar}.

Second, consider the relation between the on-shell and \msbar\
definitions of
$\ms$ and $X_t$ obtained in \eq{mstos}.  Using the integral expressions
for the loop functions that appear in \eq{fandg}, one must evaluate the
following integrals:
\be
I_n=\int_0^1\,dy\,y^n\,\ln\left[y^2(1-zx_t)+yz(x_t-2z)+z^2\right]\,,
\ee
for $n=0$, 1.  In this case, one cannot simply expand the
logarithms about $x_t=z=0$, since the integration range extends
down to $y=0$. Instead, we have used \mma\ to evaluate the
integral exactly, and then perform the double expansion in $x_t$
and $z$.  The result for $I_0$ up to ${\cal O}(x_t^2 z^4)$ is \bea
I_0 &=& -2+\pi z + z^2(2\ln z -1)-\half\pi z^3+\half z^4\nonumber
\\ &&+x_t\left[-z-z\ln z+\pi z^2+z^3(2\ln z-\half)-\half\pi
z^4\right] \nonumber \\ &&+x_t^2 \left[-\eighth\pi z-z^2(\ln z+1)+
\nicefrac{15}{16}\pi z^3+ z^4(2\ln z-\half)\right] \nonumber \\
&&+x_t^3\left[\nicefrac{1}{12}z-\nicefrac{1}{8}\pi z^2-z^3(\ln z+
\nicefrac{11}{12})+\nicefrac{15}{16}\pi z^4\right] \nonumber \\
&&+x_t^4\left[-\nicefrac{1}{128}\pi
z+\nicefrac{1}{12}z^2-\nicefrac{35}{256}\pi z^3 -z^4(\ln z
+\nicefrac{11}{12})\right] \,.  \label{expand0} \eea We have
checked the validity of this expansion using numerical
integration. One can derive a similar expression for $I_1$ either
directly, or by noting that:
\bea
I_1 &=& {1\over
2(1-zx_t)}\left[z^2(1-2\ln z)-(1-z^2)[1-\ln(1-z^2)]
-z(x_t-2z)I_0\right] \nonumber \\[6pt]
&&\qquad =-\half-z^2(\ln
z+\nicefrac{3}{2})+\pi z^3 +z^4(2\ln z-\nicefrac{3}{4})\nonumber
\\
&&\qquad \phantom{=}+x_t\left[\half z-\half\pi z^2-z^3(3\ln
z+2) +\nicefrac{9}{4}\pi z^4\right]\nonumber \\
&&\qquad
\phantom{=}+x_t^2\left[\half z^2(\ln z+2)-\nicefrac{9}{8}\pi z^3
-z^4(5\ln z+\nicefrac{11}{4})\right]\nonumber \\
&&\qquad
\phantom{=}+x_t^3\left[\nicefrac{1}{16}\pi z^2 +z^3(\ln
z+\nicefrac{19}{12})-\nicefrac{55}{32}\pi z^4\right] \nonumber \\
&&\qquad \phantom{=}+x_t^4\left[-\nicefrac{1}{24}z^2
+\nicefrac{15}{128}\pi z^3+z^4(\nicefrac{3}{2}\ln z
+\nicefrac{17}{8})\right] \,.
\label{expand1}
\eea
Inserting these
results into \eqns{mstos}{Xtos}, and expanding out the remaining
factors [{\it e.g.}, $\mgl\mt=\ms^2 z (1-z^2)^{1/2}$, {\it etc.}],
one ends up with the results given in
\eqns{eq:msmsApp}{eq:xtmsApp}.

\appendixB{Appendix~B: \boldmath{Results up to ${\cal O}
\left(\mt^4/\ms^4\right)$}}

We list the necessary formulae in order to derive the results
given in sections~3 and 4 up to terms of ${\cal
O}\left(\mt^4/\ms^4\right)$. As before, we define $x_t\equiv
X_t/\ms$ and $z\equiv \mt/\ms$. In the approximation discussed in
section~\ref{sec:diagramm}, we obtain expansions that are valid in
the limit of $z\ll 1$ and $x_t\lsim 1$.

The diagrammatic result in the on-shell scheme for the one-loop
and two-loop contributions to $\mh^2$, in the approximations used
in this paper, is given up to ${\cal O}(z^4)$ by
\bea
\mhl^{2,{\alpha}} &=&
  \frac{3}{2} \frac{G_F \sqrt{2}}{\pi^2} M_t^4 \left\{
  - 2\ln z + x_t^2\left[1-
 \nicefrac{1}{12}x_t^2 -z^2(\half-\third x_t^2) - \quarter z^4
x_t^2\right]\right\}
\label{eq:mh1ldiagosApp} \\
\mhl^{2,{\alpha\alpha_s}} &=&
  - 3 \frac{G_F \sqrt{2}}{\pi^2} \frac{\alpha_s}{\pi} M_t^4
  \Biggl\{4\ln^2 z +\Biggl[- 4 - 2 x_t^2+\nicefrac{1}{9}
z^2\left(6+42x_t+33x_t^2-26x_t^3-18x_t^4\right)\non \\
&& \qquad\qquad {}+\nicefrac{1}{9}
z^4\left(2-41x_t-10x_t^2+91x_t^3+45x_t^4\right)
  \Biggr] \ln z\non \\
&& {}  -\quarter x_t(8-x_t^3)+\nicefrac{1}{36}\pi
zx_t\left(48+24x_t-14x_t^2-7x_t^3\right) \non \\
&& {}+\nicefrac{1}{36}z^2x_t\left(60-6x_t-106x_t^2-61x_t^3\right)
-\nicefrac{1}{72}\pi z^3x_t\left(192+72
x_t-236x_t^2-111x_t^3\right) \non \\
&& {} +\nicefrac{1}{36}
z^4x_t\left(57+34x_t+95x_t^2+20x_t^3\right) \Biggr\}\,.
\label{eq:mh2ldiagosApp}
\eea
\Eq{eq:mh2ldiagosApp} is a generalization of the corresponding
formula given in \Ref{mhdiagcomp}.

The relations between the \msbar\ parameters
$\Msbar$ and $\Xtbar$ [\eq{eq:MsXtbar}] and the corresponding
on-shell parameters
$\ms^{\OS}$, $X_t^{\OS}$ can be obtained from \eqns{mstos}{Xtos} using
the expansions of \eqns{expand0}{expand1}.  The end result
up to ${\cal O}(z^4)$ is given by
\bea
\Msbarii &=& \ms^{2, \OS}\Biggl\{1 + \frac{2\alpha_s}{3\pi}
\Biggl[ -4 -\nicefrac{1}{12}z^2\left(
       12 + 24 x_t + 6 x_t^2 - 2 x_t^3 - x_t^4\right) \non \\
&& {}  +\nicefrac{1}{8}\pi z^3 x_t (16 + 8 x_t - 2 x_t^2 - x_t^3)
  +\nicefrac{1}{12}z^4 (6 - 6 x_t^2 - 23 x_t^3 - 10 x_t^4) \non \\
&& {} + \left[z^2(2-2x_t-x_t^2)+z^4x_t(5+2x_t-2x_t^2-x_t^3)\right]\ln z
      \Biggr] \Biggr\}\,,
\label{eq:msmsApp} \\
\Xtbar &=& X_t^{\OS} \frac{\mt^{\OS}}{\mt^{\MS}(\ms)}
+\frac{2\alpha_s}{3\pi} \ms \Biggl\{4-\nicefrac{1}{64}\pi z
 (128 + 64 x_t - 16 x_t^2 - 8 x_t^3 - x_t^4) \non \\
&& {}  +\nicefrac{1}{6} z^2
x_t(6 + 12 x_t + 4 x_t^2 - x_t^3)
  + \nicefrac{1}{64}\pi z^3 (128 + 32 x_t
- 128 x_t^2 - 60 x_t^3+ 17 x_t^4) \non \\[5pt]
&& {}-\nicefrac{1}{12}z^4
 (30 + 6 x_t - 6 x_t^3 - 23 x_t^4) \non \\
&& {} +\left[z^2(2 + x_t) (-2 + x_t^2)+z^4(2-5x_t^2-2x_t^3+2x_t^4)\right]\ln z
\Biggr\}\,.
\label{eq:xtmsApp}
\eea

To complete the evaluation of $\Xtbar$ we need to find a relation
between the on-shell top-quark mass, $M_t$, and the running top-quark
mass, $\mtbar(\ms)$, evaluated at the scale $\ms$.  Using \eq{mtmsos},
one obtains the following expansion:
\bea \label{mtsusymsbar}
\mtbar(\ms) &=& M_t
\left\{1 +
\frac{\alpha_s}{3\pi}\left[-4+
   6\ln z +x_t+z^2\left(\half+\quarter x_t^2+\nicefrac{1}{6}x_t^3
   % +\nicefrac{1}{12}x_t^4
   \right) \right.\right. \nonumber \\
&& \left.\left. +z^4\left(\nicefrac{1}{6}-\nicefrac{1}{24}x_t
+\nicefrac{1}{6}x_t^2+\nicefrac{1}{12}x_t^3+ \nicefrac{1}{12}x_t^4+
\nicefrac{1}{15}x_t^5\right)\right]\right\} \,.
\eea
\Eq{mtsusymsbar} was derived using DREG.  In order to obtain the
corresponding formula using DRED (which yields a formula for the
top-quark mass in the \drbar scheme), simply replace $-4$
with $-5$ in the first term of \eq{mtsusymsbar} after the left square bracket.

Note that \eq{mtsusymsbar} provides a connection between $\mtbar(\ms)$
and the on-shell mass $M_t$ in the full supersymmetric theory.
In the limit of large $\ms$ with fixed $X_t/\ms$, the
threshold correction arising from the stop mixing effects
does not vanish.  On the other hand, the \msbar\ top-quark
mass,  $\mtbar\equiv \mtsmreg(M_t)$ is defined in the low-energy
(non-supersymmetric) effective theory via \eq{eq:mtmsbar}.  Thus, in
\eq{mtsusymsbar}, we may replace $M_t$ with $\mtbar$ simply by removing
the factor of $-4\alpha_s/3\pi$.  To leading order in $\mt/\ms$, one
immediately obtains \eq{eq:mtsusy}.

%xxxxxxxxxxxxxxxxxxxxxxxxxxxxxxxxxxxxxxxxxxxxxxxxxxxxxxxxxxxxx

\vskip 2cm

%\newpage

\end{document}